\documentclass[useAMS,usenatbib]{mn2e} 
\usepackage{graphicx}
\usepackage{amsmath}
\usepackage{natbib}
\usepackage{threeparttable}
\usepackage{changes}
\usepackage{multirow}

\newcommand{\sna}{\mbox{SN~Ia}}
\newcommand{\sne}{\mbox{SNe~Ia}}
\newcommand{\mch}{$M_{\rm Ch}$}

\newcommand{\myr}{\mbox {~${\rm M_{\odot}~yr^{-1}}$}}
\newcommand{\md}{\mbox {$\dot{M}$}}      
\newcommand{\mda}{\mbox {$\dot{M}_{\rm a}$}}   
\newcommand{\porb}{\mbox {$P_{\rm orb}$}}
\newcommand{\ms}{\mbox {$M_{\odot}$}}
\newcommand{\msun}{\mbox {$M_{\odot}$}}

\newcommand{\rs}{\mbox {$R_{\odot}$}}
\newcommand{\al}{\mbox {$\alpha_{ce} \times \lambda$}}
\newcommand{\ace}{\mbox {$\alpha_{ce}$}}

\newcommand{\sws}{\mbox{SNBWDs}}

\title[Population synthesis of accreting white dwarfs]
      {Next generation population synthesis of accreting white dwarfs: 
I. Hybrid calculations using BSE + MESA}

\author[Hai-Liang Chen, T. E. Woods, L. R. Yungelson, M. Gilfanov and Zhanwen
Han]{Hai-Liang Chen$^{1,2,3,4}$\thanks{E-mail:
chenhl@mpa-garching.mpg.de}, T. E. Woods$^{4}$, L. R. Yungelson$^{5}$, M.
Gilfanov$^{4,6,7}$, Zhanwen Han$^{1,2}$\\
  $^{1}$Yunnan Observatories, Chinese Academy of Sciences, Kunming, 650011,
China\\ 
  $^{2}$Key Laboratory for the Structure and Evolution of Celestial Objects, 
Chinese Academy of Sciences, Kunming 650011, China\\
  $^{3}$University of Chinese Academy of Sciences, Beijing 100049, China\\ 
  $^{4}$Max Planck Institute for Astrophysics, Karl-Schwarzschild-Str. 1,
Garching b. M{\" u}nchen 85741, Germany\\
  $^{5}$Institute of astronomy, RAS, 48 Pyatnitskaya Str., 119017 Moscow, Russia\\
  $^{6}$Kazan Federal University, Kremlevskaya str.18, 420008, Kazan, Russia\\
  $^{7}$Space Research Institute of Russian Academy of Sciences, Profsoyuznaya
84/32,117997 Moscow, Russia\\  
}

\begin{document}

\pagerange{\pageref{firstpage}--\pageref{lastpage}} \pubyear{2013}

\maketitle

\label{firstpage}

\begin{abstract}
Accreting, nuclear-burning white dwarfs have been deemed to be candidate progenitors of type Ia supernovae, 
and to account for supersoft X-ray sources, novae, etc. depending on their accretion rates. 
We have carried out a binary population synthesis study of their populations using two algorithms. 
In the first, we use the binary population synthesis code \textsf{BSE} as a baseline for the ``rapid'' approach 
commonly used in such studies.
In the second, we employ a ``hybrid'' approach, in which we use \textsf{BSE} to generate a population of white 
dwarfs (WD) with non-degenerate companions on the verge of filling their Roche lobes.
We then follow their mass transfer phase using the detailed stellar evolution code \textsf{MESA}. 
We investigate the evolution of the number of rapidly accreting white dwarfs (RAWDs) and stably nuclear-burning 
white dwarfs (SNBWDs), and estimate the type Ia supernovae (SNe Ia) rate produced by ``single-degenerate'' systems (SD). 
We find significant differences between the two algorithms in the predicted numbers of SNBWDs at early times, and also 
in the delay time distribution (DTD) of SD SNe Ia. 
Such differences in the treatment of mass transfer may partially account for differences in the SNe Ia rate and DTD found 
by different groups. 
Adopting 100\% efficiency for helium burning, the rate of SNe Ia produced by the SD-channel in a Milky-way-like galaxy 
in our calculations is $2.0\times10^{-4}\rm{yr}^{-1}$, more than an order of magnitude below the observationally inferred value. 
In agreement with previous studies, our calculated SD DTD is inconsistent with observations. 
     
\end{abstract} 


\section{Introduction}
\label{sec:intro}

Type Ia supernovae (SNe~Ia) have been used with great success as standardizable
candles, allowing for the measurement of cosmological parameters
(\citealt{rfcc+98,pagk+99}). SNe Ia are also of great importance for galactic
chemical evolution \citep[e.g.][]{mg86}. It is widely accepted that they are 
thermonuclear explosions of carbon-oxygen (CO) white dwarfs (WDs). The compact,
degenerate structure of the exploding stars in \sne\ was recently confirmed by
early-time multiwavelength observations of SN2011fe \citep{nsct+11,bksb+12}.
However, the nature of \sne\ progenitors is still unclear (see \citealt{hkrr13}
for a recent review). The models for the progenitors of SN~Ia fall into two
categories: the single degenerate (SD) model \citep{wi73} and the double
degenerate (DD) model \citep{ty81,it84,webb84}. In the standard SD-model
a WD accretes matter from a non-degenerate companion, 
which may be a main-sequence, subgiant, or  red giant star. In order to grow, a WD must accumulate mass via nuclear-burning of hydrogen into helium, and helium into carbon and oxygen. When the WD mass
reaches \mch, the WD explodes as an SN~Ia.  

However, theoretical and observational challenges 
persist  for both scenarios. The
fundamental difficulty for the SD-model is the narrow range of accretion rates
($\sim$few$\times10^{-7}$\,\myr) for which steady nuclear-burning and efficient
accumulation of mass by the WD is possible \citep{pz78}. This requires specific
combinations of donor and accretor masses, restricting the typical delay time
between formation of a binary and a \sna\ by $\sim$1 Gyr, and similarly the peak
production of \sne\ in this channel within a similar delay time. Another problem is the treatment of the excess matter which cannot be processed
through steady nuclear-burning. This is typically \textit{assumed} either to
form an extended envelope around the WD, leading to the formation of a common
envelope, or to be lost from the system in the form of an optically thick wind.

Therefore, the viability of the SD-scenario depends critically on the 
treatment of mass transfer and resulting accretion rate, which defines
whether the WD may, presumably grow in mass. White dwarfs with different accretion 
rates are associated with different sources and phenomena, e.g. supersoft X-ray 
sources (SSSs) and novae. Comparing observations with the number of SSSs and the 
nova rate predicted by population synthesis models can be used to verify calculations, 
and also to constrain the SD-channel.

Because of the relatively high mass transfer rates needed to sustain steady
nuclear burning, these sources are almost always associated with mass transfer
on the donor's thermal timescale (thermal timescale mass transfer, TTMT).
In binary population synthesis codes, TTMT is typically accounted for using 
a simple analytic treatment. However,
such analysis typically assumes implicitly that the donor star remains in {\it
thermal equilibrium}, with the entire star (or at least its entire envelope)
responding at once, despite mass transfer being driven by the {\it thermal
disequilibrium} of the donor \citep[e.g][]{yltk95, rbf09, btn13}. 
This is particularly
important in treating mass loss from red giants -- detailed calculations reveal
that the rapid expansion of the donor envelope in response to mass transfer,
expected in the simplified treatment of adiabatic models \citep{hw87}, does not
necessarily occur \citep{wi11}. This is critical in determining the
circumstances under which a binary will undergo a common envelope (CE) phase. In
those cases where the binary will undergo a CE regardless, it is also possible
that some mass may be accreted prior to this phase, and any
accreting WD may appear briefly as an SSS. This is unaccounted for in the
traditional treatment of mass transfer in population synthesis.

In this paper (Paper I), we investigate in detail mass transfer in the semidetached systems with nuclear-burning WD (NBWD) accretors and main-sequence, Hertzsprung gap and red-giant donors.  
We pay special attention to the systems in which WDs burn hydrogen steadily (SNBWDs) and to the systems with accretion rates exceeding the upper limit for steady burning, but too low for the formation of a common envelope, 
\citep[``rapidly accreting white dwarfs'' (RAWDs),][]{lv13} \footnote{It was shown by 
\citet{1971AcA....21..417P} that putting a $\simeq 10^{-3}$\,\ms\ hydrogen-helium envelope atop a hot
($\log T_e=5.0$) carbon-oxygen WD transforms it into a red giant ($\log T_e=3.6$); 
this may be avoided, if excess of the matter is removed by postulated optically-thick stellar wind 
\citep{hkn99}.}.
For this, we produce a grid of $\sim3\times10^{4}$ evolutionary sequences of close binary models with different initial combinations of WD accretors and nondegenerate donors, and with differing orbital periods at the onset of Roche lobe overflow, 
calculated by the detailed stellar evolutionary code \textsf{MESA} \citep{pbdh+11,pcab+13}.
Our models are compared with the ones obtained using analytic descriptions of mass-transfer. 
In order to relate our work to observations, we compare the predicted evolution
of the numbers of SNBWD, RAWD, and the rates of \sne\ given two star formation
histories: an instantaneous burst of star formation, and a constant star formation rate for 
10\,Gyr, approximating early and late type galaxies respectively. In a subsequent paper (hereafter Paper II), we will incorporate spectral models for nuclear-burning white dwarfs. This will allow us to more meaningfully test the predictions of our model.

We describe the method of calculations in \S\ref{sec:method}, highlight the
effect of varying treatments of TTMT in $\S$\ref{TTMT}, follow with a
discussion of how some observables vary with changing MT treatment in
\S\ref{sec:res}, in particular the predicted populations of RAWDs, SNBWDs, and
\sne. Summary and conclusions are presented in \S\ref{sec:sum}. 
 
\section{The method of calculations}
\label{sec:method}

\subsection{Mass loss treatment in binary population synthesis}
\label{sec:mlt}

The method applied to study different populations of binary stars and the products of their evolution is binary population synthesis (BPS). 
In population synthesis, one convolves the statistical data on initial parameters and birthrates of binaries with scenarios for their evolution.  
This allows one to study birthrates and numbers of binaries of different classes and their distributions over observable parameters.

There are two basic algorithms applied to study semidetached stages of evolution in BPS codes. The ``rapid'' one employs analytic formulae, approximating each evolutionary phase using simple fits from detailed calculations. Mass transfer is accounted for by calculating the radial response of the donor star and its Roche radius. Alternatively, one may employ a ``hybrid'' approach which entails two steps. 
Relevant to our present study, first we obtain the population of WD binaries with nondegenerate donors at the onset of mass transfer by means of a BPS code. Here we use the publicly available code \textsf{BSE}\footnote{http://astronomy.swin.edu.au/~jhurley/bsedload.html} \citep{htp02}, which we have modified slightly (see below). 
In the second step, in order to obtain an accurate description of post-RLOF mass-loss rates, we compute the mass transfer rate and response of the donors in this population by drawing from a grid of $3\times10^{4}$ evolutionary sequences of models for WDs with MS, HG or FGB companions computed by \textsf{MESA} \citep{pbdh+11,pcab+13}, in practice, using about $\sim$4000 such tracks. The advantage of this approach is the possibility to describe \md\ accounting for the response of the donor. This also allows one to avoid the exclusion of any short evolutionary stages.

It is known from the earliest studies of close interacting binaries 
\citep[see, e.g. ][]{1960ApJ...132..146M,pzz69,ps72a} that, depending on the evolutionary status of the 
Roche-lobe-overflowing star (the donor) and the mass ratio of the components, the 
donor may lose mass on a timescale defined by the dynamical, thermal, or nuclear 
evolution of the donor, or the loss of angular momentum. In practice, this means that the
mass-loss rate depends on relations between the response of the Roche lobe radius to mass loss
$\zeta_{RL}\equiv\left(\frac{\partial \ln R_{RL}}{\partial \ln M_1}\right)$,
the adiabatic hydrostatic response of the stellar radius 
$\zeta_{ad}\equiv\left(\frac{\partial \ln R}{\partial \ln M_1}\right)_{ad}$,
the thermal-equilibrium response of the same
$\zeta_{th}\equiv\left(\frac{\partial \ln R}{\partial \ln M_1}\right)_{th}$,
the nuclear evolution of the radius, and finally the angular momentum loss timescale. 
If  $\zeta_{ad} > \zeta_{RL} > \zeta_{th}$, the star remains in hydrostatic equilibrium, 
but does not remain in thermal equilibrium; in this case mass loss occurs on the thermal 
timescale of the star.
If  $\zeta_{RL} > \zeta_{ad}$, the star cannot remain in hydrostatic equilibrium, and 
mass loss proceeds on the dynamical timescale.
If $\zeta_{ad}, \zeta_{th} > \zeta_{RL}$, mass loss occurs due to the expansion of the star during its evolution on the nuclear timescale, or due to the shrinkage of the Roche lobe owed to angular momentum losses. 

Dynamical or thermal timescale mass loss is common for initial stages of mass exchange. 
In ``rapid'' BPS codes it is assumed that, if RLOF leads to dynamical mass loss (according to 
some assumed criteria), then the formation of a common envelope is unavoidable. In this case, no further computations of the mass transfer rate or the response of the donor star are carried out.  
Mass loss is assumed to occur on the thermal timescale if, after removal of an 
infinitesimally small amount of mass, the radius of the star in thermal 
equilibrium is predicted to be
larger than the (volume-averaged) Roche lobe radius.
In the simplest formulation, the mass loss rate is approximated as 
\begin{equation}
\label{eq:mdot_th}
\dot{M}_{\rm th}=M/\tau,
\end{equation}
where $M$ is the mass of the Roche -lobe filling star and $\tau$ is an estimate 
of the thermal timescale. In general, the definition of $\dot{M}_{\rm th}$ is not unique and differs between codes because of our present lack of understanding of what fraction of the star is involved in mass exchange (i.e. to what depth to we evaluate $\tau _{\rm{th}}$?). 
Then, typically, one writes $\dot{M}_{\rm th} = k \cdot R_{\rm   d}L_{\rm d}/GM^{\prime}$, where, in the simplest cases, $k\sim$1, $M^{\prime}$ is the instantaneous total mass of the star or the mass of the stellar envelope.  
In slightly more sophisticated cases, $k$ may be a function of the mass-radius response functions of the donor and its Roche radius \citep[e.g.][]{Ivanova04}.
Moreover, for $R_d$\ and $L_d$\, the values corresponding to stars in {\bf thermal equilibrium} are usually taken, despite mass transfer being driven by the {\bf thermal disequilibrium} of the donor star.  
As a result, just after RLOF $\md$ is constant or slowly declining as $M_d$ falls (see Fig.~\ref{fig:examples} below).
However, \citet{ldwh00} (see also \citealt{prp02}) clearly showed that equations similar to  
Eq.~(\ref{eq:mdot_th}) provide only the order of magnitude of the mass transfer rate.

Although detailed calculations accounting for the response of the donor are preferable, 
the use of the ``hybrid'' technique has so far been limited by the available computing power. 
If too crude a grid of evolutionary models is used, it may result in spurious effects. 
Only recently, with the advent of the rapid and robust 
stellar evolutionary code \textsf{MESA}, 
has it become possible to increase the number of sufficiently detailed
tracks by an order of magnitude, to 
several 10,\,000, which can be used to adequately describe the evolution of a parental 
population of $\sim 100,\,000$ systems.

Earlier, grids of detailed evolutionary tracks leading to the systems or events of desired type 
were combined with detailed BPS codes, e.g., by \citet{prp03} for low and intermediate mass X-ray binaries,
\citet{hp04} for SNe Ia in semidetached systems, \citet{mrpn08} for ultraluminous X-ray sources. Advantages of this
method were recently discussed by \citet{2012JPhCS.341a2008N}. In this paper, we present the first attempt to use
grids of tracks computed by \textsf{MESA} to investigate accreting WDs.

\subsection{Binary population synthesis for NBWDs}
\label{sec:bps}

In this paper, we employ both the ``hybrid'' approach and the ``rapid'' approach described above. In either case, each evolutionary track for every binary must be scaled by a relative ``weight'' which accounts for the total number of binaries the track represents. 
To evaluate the relative ``weight'' of a binary, we take the IMF of primaries from \citet{krou01} 
with lower and upper mass cutoffs $m_{L} = 0.1\,M_{\odot}$ and $m_{U} = 100 M_{\odot}$.  
We adopt a flat mass ratio distribution \citep{1979SvA....23..290K}.
Binary separations in the range between $10R_{\odot}$ \ and $10^{6}R_{\odot}$ are drawn from a flat distribution in logarithmic space \citep{abt83}. 
We assume 50\%-binarity, i.e., 2/3 of stars are components of binary systems \citep{Duquennoy_Mayor91}. 

Using the latest version of \textsf{BSE}, we computed the evolution of binaries for a regular grid of stars with primary masses varying from 0.9\ms\ to 12\ms\ in logarithmic steps of 0.0125, mass ratio of components between 0 and 1 in steps of $\Delta q$ = 0.0125, and separations of components from 10\,\rs\ to $10^4$\,\rs\ in logarithmic steps of 0.025. Altogether, this gives us a grid of $864,000$ systems. 
As a subset of this grid, we obtain the total population of WD binaries with nondegenerate donors at the onset of mass transfer.

\begin{figure}
\centering
\includegraphics[height=0.25\textheight]{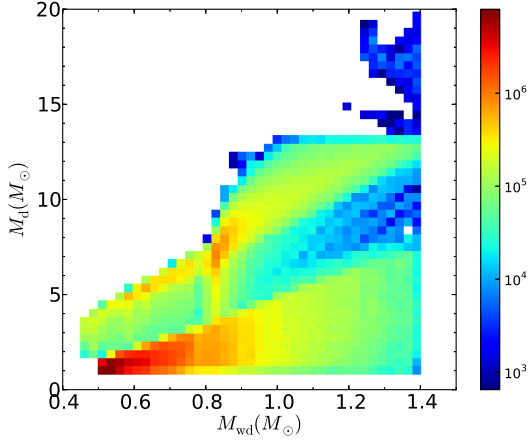}
\caption{WD and donor masses distribution for the population of
  WD+(nondegenerate companion) binaries with different orbital periods
  at the onset of mass transfer for a $10^{11} M_{_{\odot}}$ galaxy in the model B1+M (see table 2).}  
\label{fig:progen}
\end{figure}

The WD masses at the onset of mass transfer range from 0.50\,\ms\ to 1.40\,\ms\ and companion masses -- from 0.8\,\ms\ to 20\,\ms\ (see Fig.~\ref{fig:progen}), which are consistent with other binary population synthesis studies (see Fig 3. and 4. in \citealt{tcmr13}).  
For the systems with primary masses or donor masses $\leq$1.4\,\ms\ we replaced the original \textsf{BSE} implementation of magnetic braking by the prescription suggested by \citet[][ Eq.~(34) with $\gamma=3$]{rvj83}; this is identical to the magnetic braking law implemented in \textsf{MESA}.

To describe common envelope evolution, we use the prescription suggested by \citet{webb84}, with the inclusion of the ``binding energy parameter'' \citep{dek90}:

\begin{equation}
  \alpha_{ce}(\frac{Gm_{\rm{d,f}}m_{\rm a}}{2a_{\rm f}}-
  \frac{Gm_{\rm d,i}m_{\rm a}}{2a_{\rm i}}) 
  = \frac{Gm_{\rm d,i}m_{\rm d,e}}{\lambda R_{\rm d,r}}.
\label{eq:ce}
\end{equation}

\noindent Here $m_{\rm d,i}$ and $m_{\rm d,f}$ are initial and final donor mass, respectively, $m_{\rm d,e}$ is the donor envelope mass, $m_{a}$ is the accretor mass, $a_{\rm i}$ and $a_{\rm f}$ are the initial and final binary separations, $R_{\rm d,r}$ is the Roche lobe radius of the donor at the onset of mass transfer, $\ace$ is the fraction of the orbital energy used to eject the common envelope and $\lambda$ is a parameter which characterizes the binding energy of the donor's envelope.  
The long-standing problem of the formalism given by Eq.~(\ref{eq:ce}) is that $a_f/a_i$ depends on the product of $\al$ and these two still uncertain parameters cannot be separated \citep[see ][ for the latest detailed discussion]{ivanova_al_ce13}. 
It is evident that $\lambda$ should not be constant along the evolutionary track of a star
\citep[e.g][]{dt00}. It remains uncertain whether a fraction of the binding energy of the donor may contribute to expelling 
the envelope and whether there are any other sources contributing to this process.

\citet{dkw10} computed $\lambda$'s for 1 - 8\,\ms\ AGB stars with varying radii, postulating that the
core-envelope interface is located at the position within the star with hydrogen abundance 
$X_c=0.1$. Then they found that, assuming $\alpha_{\rm CE} > 0.1$,  it is possible to account for the population of post common envelope binaries found by the SDSS. 
A similar approach led \citet{zsgn10} to
constrain the common envelope efficiency to  $0.2 < \alpha_{\rm CE} < 0.3$\footnote{
Note, their sample of post-CE binaries consisted of WD with M-type main-sequence companions. 
For more massive companions the energetics of the CE phase may differ, see the quoted 
paper and \citet{zsp14}.} 
\citet{rt12}, based on results of 3D hydrodynamical
calculations, limited $\alpha_{\rm CE}$ from above by 0.4 - 0.5.

Having these uncertainties in mind, we produced a set of WD+(nondegenerate companion) models using \textsf{BSE}, with all CE events following Eq.~(\ref{eq:ce}) assuming a constant \al=0.25. 
In another set of computations, we used fitting formulae for $\lambda$\ \citep{lvk11} and \ace=0.25. 

 In the MESA grid, WD masses range from $0.5 M_{\odot}$ to $1.3 M_{\odot}$ with an interval of $0.1 M_{\odot}$, with an additional final step at $1.35M_{\odot}$.  
The donor masses range from $0.9 M_{\odot}$ to $2.5 M_{\odot}$ with an interval of $0.025 M_{\odot}$, with an interval of $0.1$\,\ms\ from $ 2.6 M_{\odot}$ to $3.5 M_{\odot}$, 
with an interval of 0.50 from 4.0\,\ms to 10.0\,\ms, and with an interval of 1.0\,\ms\ from 10.0\,\ms\ to 15\,\ms.  
Initial orbital periods $\mathrm{log}(P_{\mathrm{orb}}/{\rm day})$ cover the range  from $-0.3$ to $2.9$ with a logarithmic 
step of $0.1$.  
If the parameters of a binary produced in the first step are out of this grid, we computed it individually. In the second step, for every binary 
we choose the nearest track in grid of \textsf{MESA} calculations in order to follow the evolution of the system.

The tracks are computed for typical Population~I composition with initial hydrogen abundance X = 0.70, helium abundance Y = 0.28 and metallicity Z = 0.02.

The retention efficiency of matter accumulated by WDs was estimated on the basis of several ``critical'' accretion rates. 
Accreted hydrogen burns stably if $\dot{M}_{\rm cr} \leq \mda \leq \dot{M}_{\rm max}$, where, we employ the following approximations to the results of \citet{it89} 
\begin{multline}
\log(\dot{M}_{\rm max}) \approx -4.6 \times {\rm M^4_{WD}}+ \\
 17.9  \times {\rm M^3_{WD}}-26.0\times {\rm M^2_{WD}}+17.5 \times {\rm M_{WD}}-11.1,
\end{multline}
\begin{equation}
\log(\dot{M}_{\rm cr}) \approx -1.4 \times {\rm M}^2_{WD} + 4.1 \times {\rm M_{WD}} -9.3.
\end{equation}
Masses are in units of \ms\ and rates in units of \myr.
For  $\mda  >10^{-4}$\,\myr, an optically thick wind can not be sustained \citep{hkn99} and a CE is formed.
If $\dot{M}_{\rm max} < \mda  \leq 10^{-4}$\,\myr, the excess of unburned matter is isotropically reemitted from the system by an optically thick wind. 
If $\dot{M}_{\rm cr}> \mda\ \geq 10^{-8}$\,\myr, H burns in mild flashes.
Below $10^{-8}$\,\myr\ burning flashes are strong and may even erode the dwarf. We apply 
in this regime a  fitting formula for the mass retention efficiency $\eta_{\rm H}$\ based on the results of \citet{pk95} and
\citet{ypsk05}:
\begin{equation}
\label{eq_ef_fit}
\eta_{\rm H}=-0.075\times(\log(\dot{M}_{\rm a}))^2-\\
1.21\times \log(\dot{M}_{\rm a})-4.95.
\end{equation}
Note, $\eta_{\rm H}$ is based on the results for WD temperature $\rm{T_{WD}} = 3
\times10^{7}K$. We neglect weak dependence of  $\eta_{\rm H}$ on WD mass.

To summarize (rates are in \myr): 
\begin{equation}
\eta_{\rm H}=\begin{cases} 
\mbox{CE} & \dot{M}_{\rm a} > 10^{-4}\\
\dot{M}_{\rm max}/\dot{M}_{\rm a}  & \dot{M}_{\rm max} < \dot{M}_ a \le
10^{-4} \\
1.0                       & \dot{M}_{\rm cr} \le \dot{M}_a \le
\dot{M}_{\rm max}\\
{\rm  linear\ interpolation} & \\
{\rm between\ 1\ and\ Eq.~(\ref{eq_ef_fit})} & 10^{-8} < \dot{M}_a < \dot{M}_{\rm
cr}\\
{\rm Eq.~(\ref{eq_ef_fit})} & 10^{-12} \le \dot{M}_{\rm a} \le 10^{-8}.
\end{cases}
\end{equation}

The ranges of WD masses and accretion rates of steady or unsteady burning regimes of H and He are not identical \citep[see, e.g., ][]{it89}. 
For this reason, in population synthesis studies retention efficiency of He,  $\eta_{\rm He}$, is considered as a function of $M_{\rm WD}$ and the rate of H-accumulation. Then, total mass accumulation efficiency is taken as $\eta_{\rm H} \times \eta_{\rm He}$.
The value of $\eta_{\rm He}$ is especially uncertain \citep{btn13,iss13,nst13,wbbp13}.
We assumed, for simplicity, $\eta_{\rm He}=1$. This provides a strict upper limit on the SNe Ia rate.

In the ``rapid'' approach, the following mass-ratio based criteria for the formation of a CE were applied: $q = 3$ for MS donors, $q = 4$ for HG donors.
These values are supported by detailed binary evolution studies \citep{hp04,wlh10}. If the donor stars are on the FGB and AGB,
we use the prescription of \cite{hw87}, despite the fact that this underestimates the stability of mass transfering red giants \citep{wi11}, for lack of an alternative prescription. 
A summary of the NBWDs population models computed in this paper is presented in Table~\ref{tab:runs}. 
The use of four different models allow us to compare our results for differing treatment of the second mass transfer phase (analytic or MESA) with that of our choice in CE parameterization (fixed $\alpha\lambda$, or an analytic fit 
of $\lambda$ from detailed evolutionary calculations with fixed $\alpha$).

\begin{table}
\caption{Computed models} 
\label{tab:runs}
 \begin{threeparttable}
  \begin{tabular}{ccccc}
  \hline 
  model & $\lambda$       &  $\alpha$    & code\\
  \hline
  B1    & fit\tnote{a} &  0.25      &  \textsf{BSE} only     \\
  \hline
  B1+M    & fit\tnote{a} &  0.25      &    \textsf{BSE + MESA}    \\
  \hline
  B2     & \multicolumn{2}{c}{$\alpha\times \lambda = 0.25$} &  \textsf{BSE} only\\
  \hline
  B2+M     & \multicolumn{2}{c}{$\alpha\times \lambda = 0.25$} &  \textsf{BSE + MESA}\\
  \hline
  \end{tabular}
  \begin{tablenotes}
  \item[a]{The fitting formula from \cite{lvk11} for $Z = Z_{\odot}$.}
  \end{tablenotes}    
 \end{threeparttable}
\end{table}

\section{Comparison of mass transfer treatments}
\label{TTMT}

\begin{figure*}
\centering
\includegraphics[height=0.44\textheight]{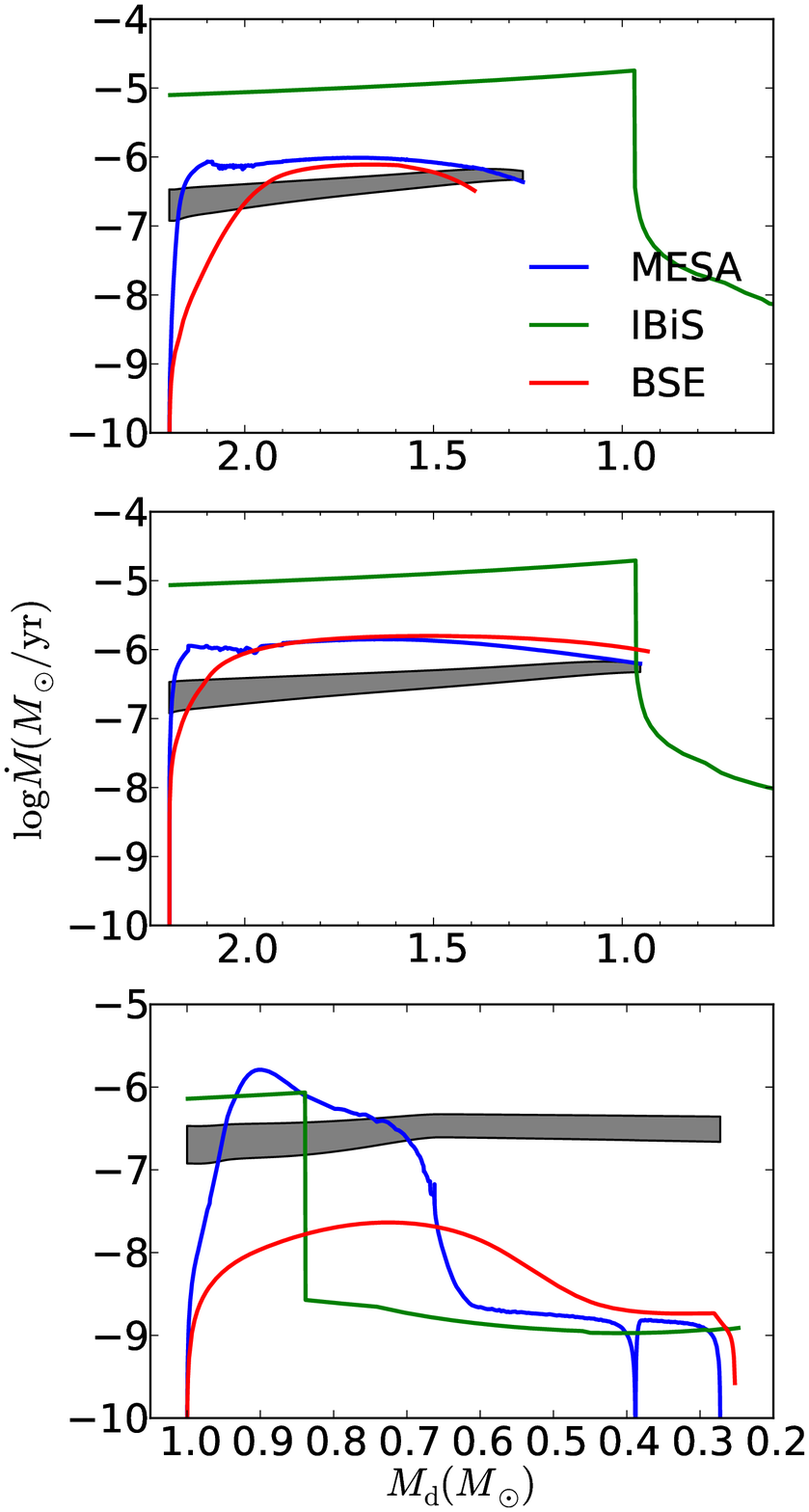}
\includegraphics[height=0.44\textheight]{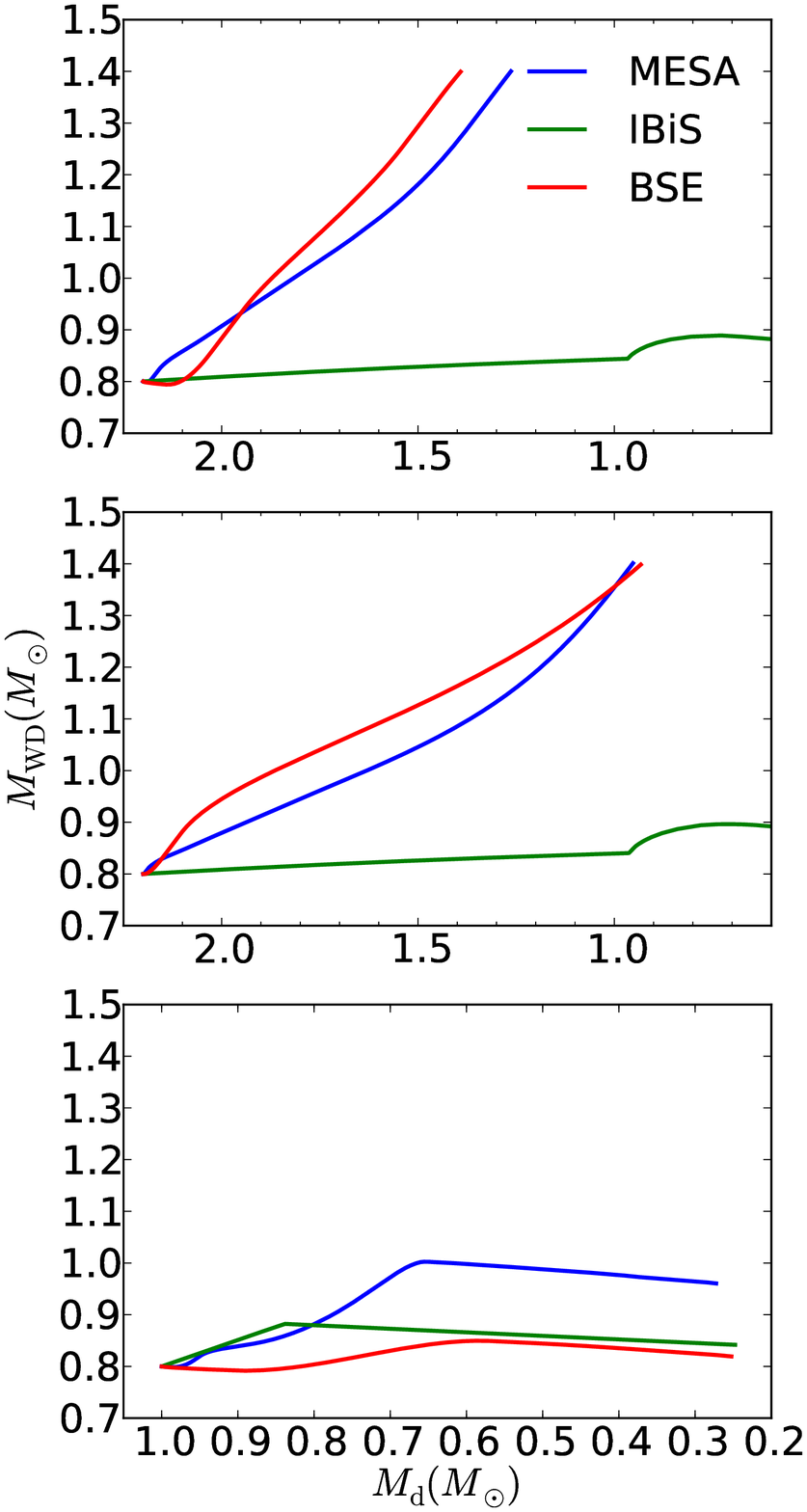}
\includegraphics[height=0.44\textheight]{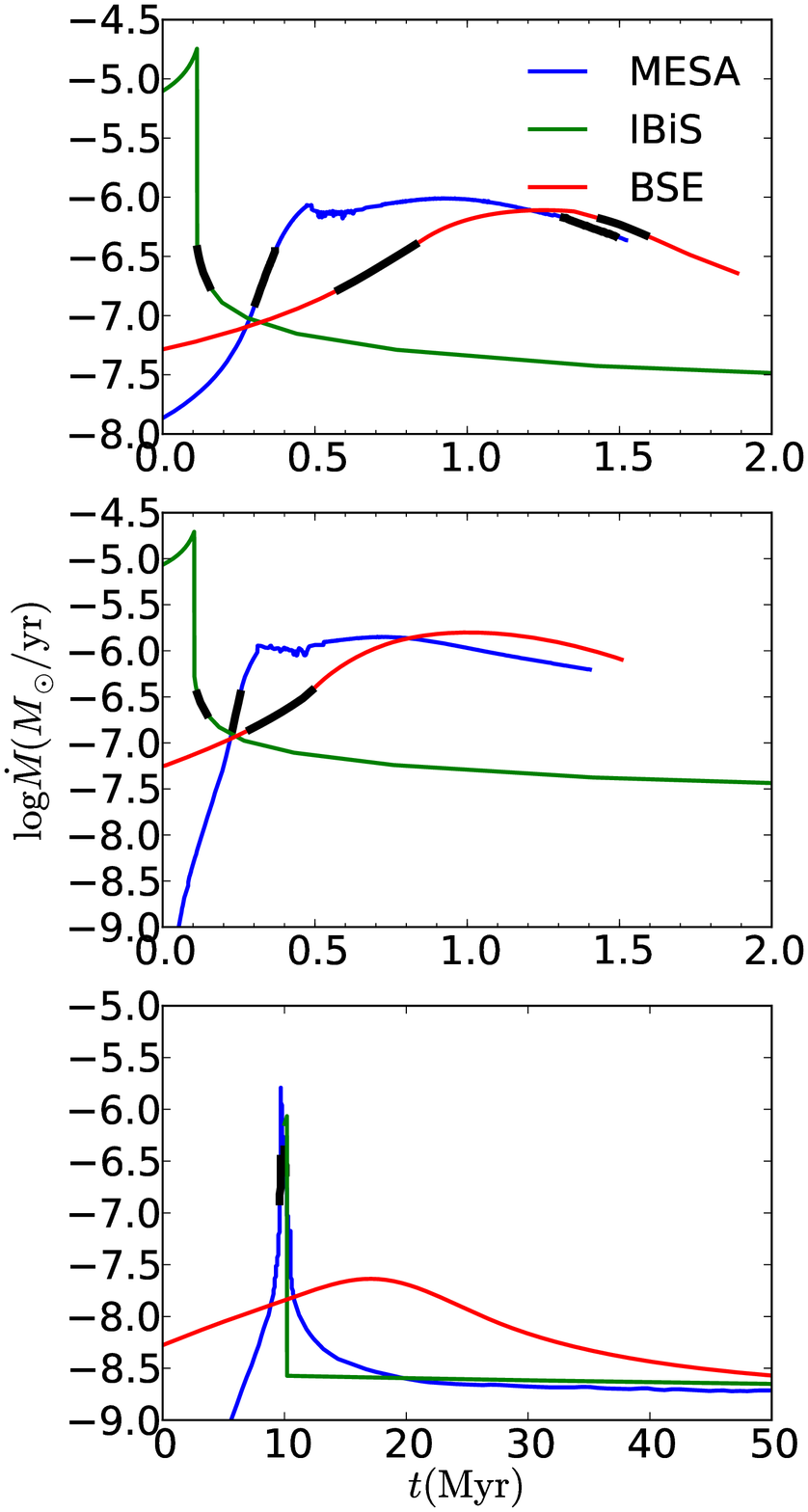}
\caption{ Comparison of the evolution of mass transfer rate and mass
  of the accretor as a function of donor mass (left and middle
  panels). The right set shows the dependence of \md\ on time.  At the
  onset of mass transfer $M_{\rm WD}$ = 0.80\,\ms, $M_{\rm d}$ =
  2.20\,\ms,  \porb = 0.80, 2.0 days in the upper and
  middle panels, respectively. In the lower panel, the binary
  parameters are $M_{\rm WD}$ = 0.8\,\ms, $M_{\rm d}$ = 1.00\,\ms, 
  \porb = 3.0 days. For these three binaries, mass
  transfer begins on the MS, HG and RG branch, respectively.  In the right set, the 
  thick black line shows the time spent in the stable burning regime. }

\label{fig:examples}
\end{figure*}

Here we present examples of computations using two simplified prescriptions for mass transfer and compare them with the results of our detailed stellar evolution calculations using \textsf{MESA}.
First, we apply the prescription used in \textsf{BSE}. In solar units,
the ``thermal timescale'' mass transfer rate is  defined as 
$10^{-7} M_{\rm d,0}RL/M^{\prime}$~\myr, where 
$M^{\prime}=M_{\rm d}-M_c$, $M_c$ is the mass of stellar
core. This mass transfer rate is compared to the ``nuclear timescale'' mass transfer rate
given by an \textit{ad hoc} formula
$\md=3\times10^{-6}\left[\min(M_{\rm d},5.0)\right]^2\left[\ln(R/R_{\rm RL})\right]^3$~\myr.
Then, the minimum of two values for \md\ is chosen.  In the illustrative 
cases presented below, \md\, corresponds to the first of these two formulae. 

As the second prescription, we use the formulation applied in the code \textsf{IBiS} 
\citep[e. g. ][]{1998ApJ...497..168Y}.
If $\zeta_{th}<\zeta_{\rm RL}$, $\md=3.15\times10^{-7}RL/M_{\rm d}^2$~\myr.
Otherwise, if mass transfer is thermally stable, \md\ is defined by the timescale of growth of the degenerate He-core of the donor.
The stabilizing effect of an optically thick wind of from the WD is taken into account in the high-$\dot M$ regime. 

Finally, in the detailed stellar evolutionary code \textsf{MESA}, \md\ is computed
as a stationary subsonic isothermal flow through the vicinity of $L_1$, with an additional 
assumption that $(R-R_{\rm RL})/H_p \approx1$, folllowing 
\citet{1988A&A...202...93R}. Here $H_p$ is the pressure scale height. 

In Fig.~\ref{fig:examples}, we show example tracks for three cases of binary evolution, where in each case we have either used one of the two simplified algorithms described above, or \textsf{MESA}. 
In the upper and middle rows of Fig. ~\ref{fig:examples}, 
 the initial white dwarf mass is $M_{\rm WD} =
0.80 M_{\odot}$, the initial donor mass is $M_{\mathrm d}=2.2M_{\odot}$, and 
the initial orbital periods are $P_{\rm  orb} = $ 0.80 (upper) \& 2.20 days (middle row). 
In the lower set of 
panels, $M_{\rm WD} = 0.80 M_{\odot}$, $M_{\rm d}$ =
1.00 $M_{\odot}$, $P_{\rm orb}$ = 3.0 days. 
In these three cases, the binaries begin mass transfer on the MS, HG and FGB,
respectively. In the initial stage of mass transfer in all three systems
$\zeta_{th}<\zeta_{\rm RL}$ and mass transfer should  proceed on thermal timescale. 
These binaries should not be considered representative of the total population; rather, we 
aim here to represent those systems in which a WD reaches \mch. Note in particular that WD binaries in which the companion overflows its Roche lobe on the late main-sequence or Hertzsprung gap constitute the overwhelming majority of ``successful'' progenitors of \sne\ 
in all sets of our calculations. 

Note that the binary illustrated in the lower row of Fig. ~\ref{fig:examples} would, upon overfilling its
Roche lobe on the RGB, form a  CE if the \citet{hw87} criterion for stability of mass-transfer were applied (for the purposes of comparison we ignore this criterion here).
Although this criterion is widely used, it may overestimate the number of systems which undergo unstable mass-transfer  \citep{wi11,php12}.  In the RG case, the binary has short thermal timescale mass transfer phases in the \textsf{IBiS} and
\textsf{MESA}-based calculations. 
Apparently, the formal choice of lower mass transfer mass rate in \textsf{BSE}, ignores existence of TTMT-stage in this case. 

In table ~\ref{tab:mtcom}, we present the duration of the RAWD and SNBWD phases, as well as the mass lost by the donor, and the mass accreted in each phase, for each example. \\

 It is worth noting that, 
with different mass transfer prescriptions, the mass accreted by the WD and the duration of the RAWD and
SNBWD phases in different codes are quite different. 
Compared to the detailed stellar evolution calculations, \textsf{BSE} usually 
overestimates the duration of the SNBWD phase and underestimates the RAWD phase. The \textsf{IBiS} code underestimates both of them in the MS and HG donor cases. 
In the MS and HG cases, the amounts of mass accreted during RAWD phase are comparable in \textsf{BSE} and \textsf{MESA} calculations, while they are an order of magnitude smaller in \textsf{IBiS} calculations. In the RG case, $\Delta \rm{M}_{\rm{WD}}$ during the RAWD phase
is comparable in \textsf{IBiS} calcultions and in \textsf{MESA} calculations.
In addition, in all three 
cases, the amount of mass
accreted during the RAWD phase is larger than that in the SNBWD phase.This points again 
to the need for a proper understanding of accretion at rates exceeding $\dot{M}_{\rm max}$. 
The primary reason for the difference between \textsf{IBiS} calculations and \textsf{BSE} or \textsf{MESA} calculations is that the duration of  the phase of effective
mass-accumulation  ($\dot{M}_{\rm a} > \dot{M}_{\rm cr}$) in \textsf{IBiS}  is 
about one order of magnitude smaller than that in \textsf{BSE} and \textsf{MESA}.

The difference between \textsf{IBiS} and \textsf{BSE}
calculations is partially due to the fact that approximations for \md\ in \textsf{IBiS} are  based on results of calculations for binary stars,  while in 
\textsf{BSE} approximations are based on computations 
for single stars. Both of them produce results different from those,
obtained using \textsf{MESA}. 
Evidently, the difference in mass transfer rate behaviour
has an important impact on the results of binary population synthesis for 
NBWDs.  Below, we will show this by using different BPS algorithms.

\begin{table*}
\caption{Comparison of the duration of RAWD and SNBWD phases, accreted mass $\Delta \rm{M}_{\rm{WD}}$ in RAWD and SNBWD phases, mass lost by the donors for the three examples shown in 
Fig.~\ref{fig:examples}.
Note that these numbers do not represent the typical values in the population, which will be addressed in a subsequent paper.}
\label{tab:mtcom}

\begin{tabular}{cccccccccccc}
\hline
\multirow{3}{*}{Example} & \multicolumn{3}{c}{MS donor} &\multicolumn{3}{c}{HG donor} & \multicolumn{3}{c}{RG donor} \\
\cline{2-10}
&BSE&IBiS&MESA&BSE&IBiS&MESA &BSE&IBiS&MESA\\
\hline
Duration of RAWD phase  (Myr)&0.576 & 0.114& 0.954&0.926 & 0.108& 1.14& 0.0& 0.204& 0.268\\
\hline
Duration of SNBWD phase (Myr)&0.397 & 0.040& 0.233& 0.201& 0.035& 0.039& 0.0& 0.0& 0.196\\
\hline
$\Delta \rm{M}_{\rm{WD}}$ in SNBWD phase ($M_{\odot}$)& 0.1516& 0.0089& 0.1068& 0.049& 0.0087& 0.010& 0.0& 0.0& 0.059\\
\hline
$\Delta \rm{M}_{\rm{WD}}$ in RAWD phase ($M_{\odot}$)& 0.3088& 0.0435 & 0.4723& 0.5480& 0.0417& 0.5885& 0.0& 0.0816& 0.0997\\
\hline
$\Delta \rm{M}_{\rm{d,ml}}$ of donor star ($M_{\odot}$)& 0.8498& 1.8860& 0.93674& 1.3264& 1.8728& 1.2475& 0.7476& 0.7515& 0.7276\\
\hline
\end{tabular}
\end{table*}    

\section{Results and Discussion}
\label{sec:res}

\subsection{Population synthesis of accreting WDs}
We model two cases of star formation: \\
(I) A starburst --- stellar population with mass $M_{\rm t} =10^{11}
M_{\odot}$
is formed at $t = 0$. \\
(II) Constant star formation rate --- a galaxy has a constant star
formation rate of $1 M_{\odot}/\rm{yr}$\ for 10\,Gyr. 

\subsubsection{The Number of SNBWD}

\begin{figure}  
\centering
\includegraphics[width=84mm]{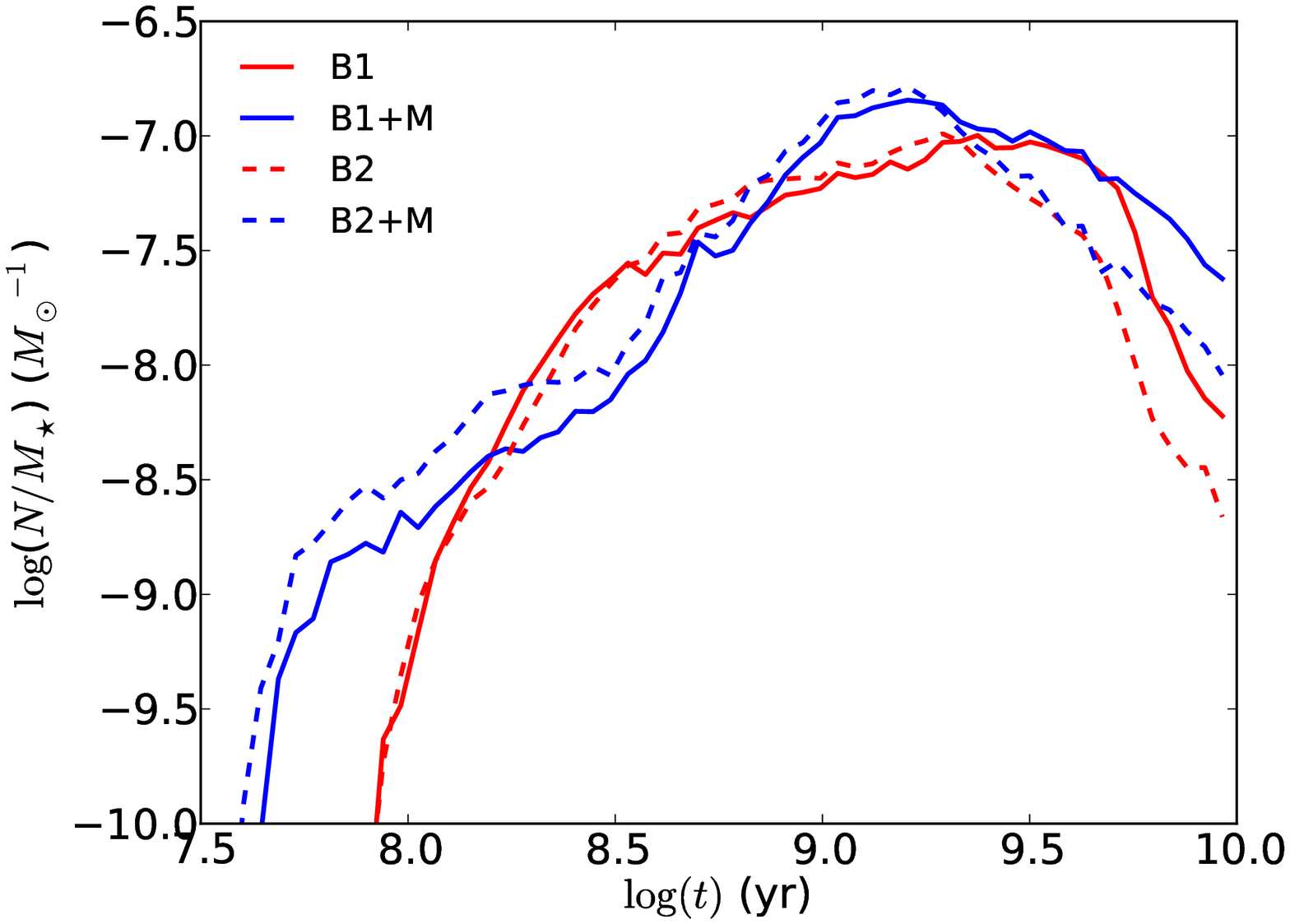}
\includegraphics[width=84mm]{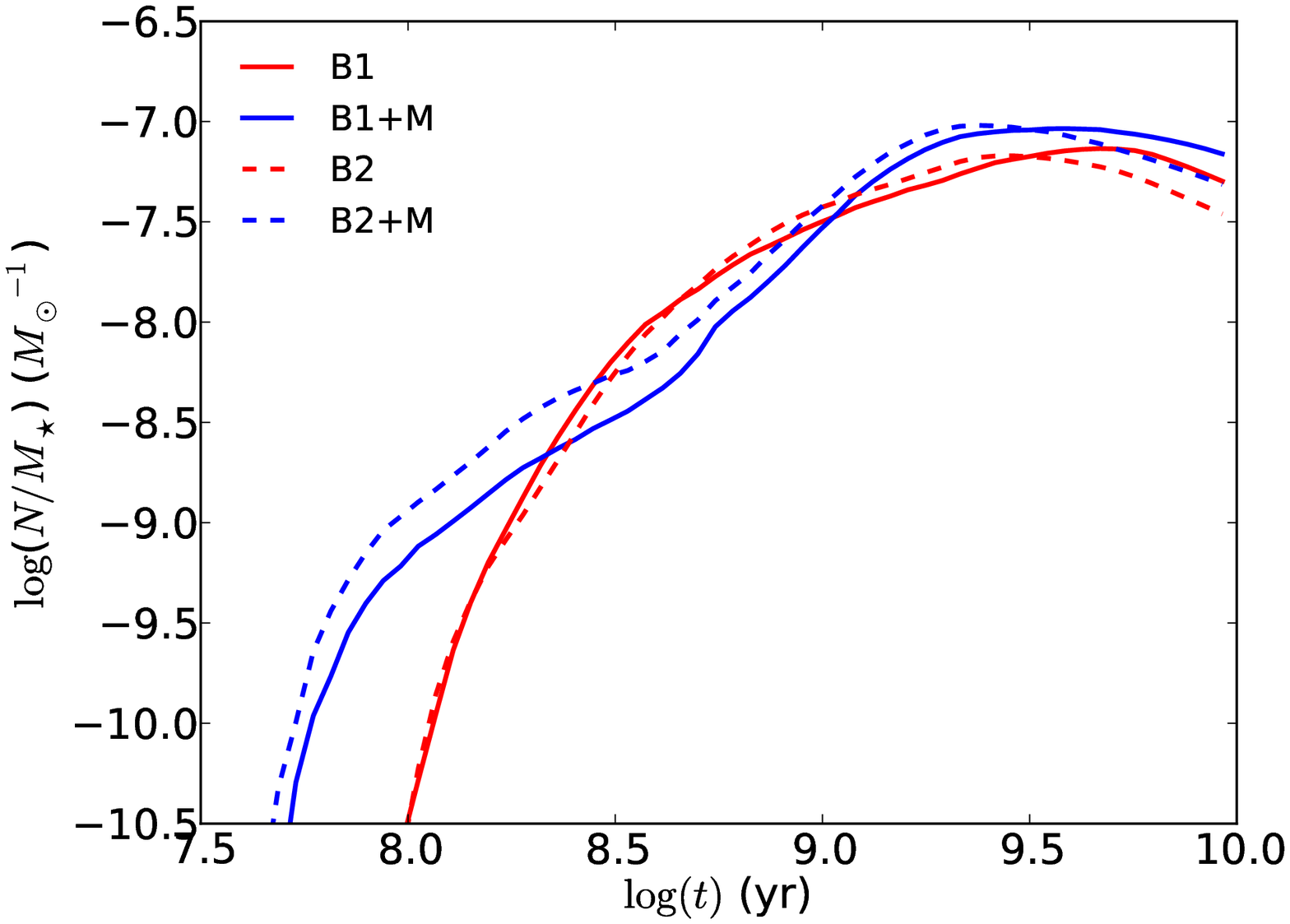}
\caption{The number of SNBWDs normalised to the total stellar mass
  for starburst case (upper panel) and constant SFR
  case with SFR = $1\,M_{\odot}/$yr (lower panel). The blue and red lines show the results computed
  with \textsf{BSE+MESA} and \textsf{BSE} only, respectively.  
}
\label{fig:numsss}
\end{figure}

\begin{figure}   
\centering
\includegraphics[width=84mm]{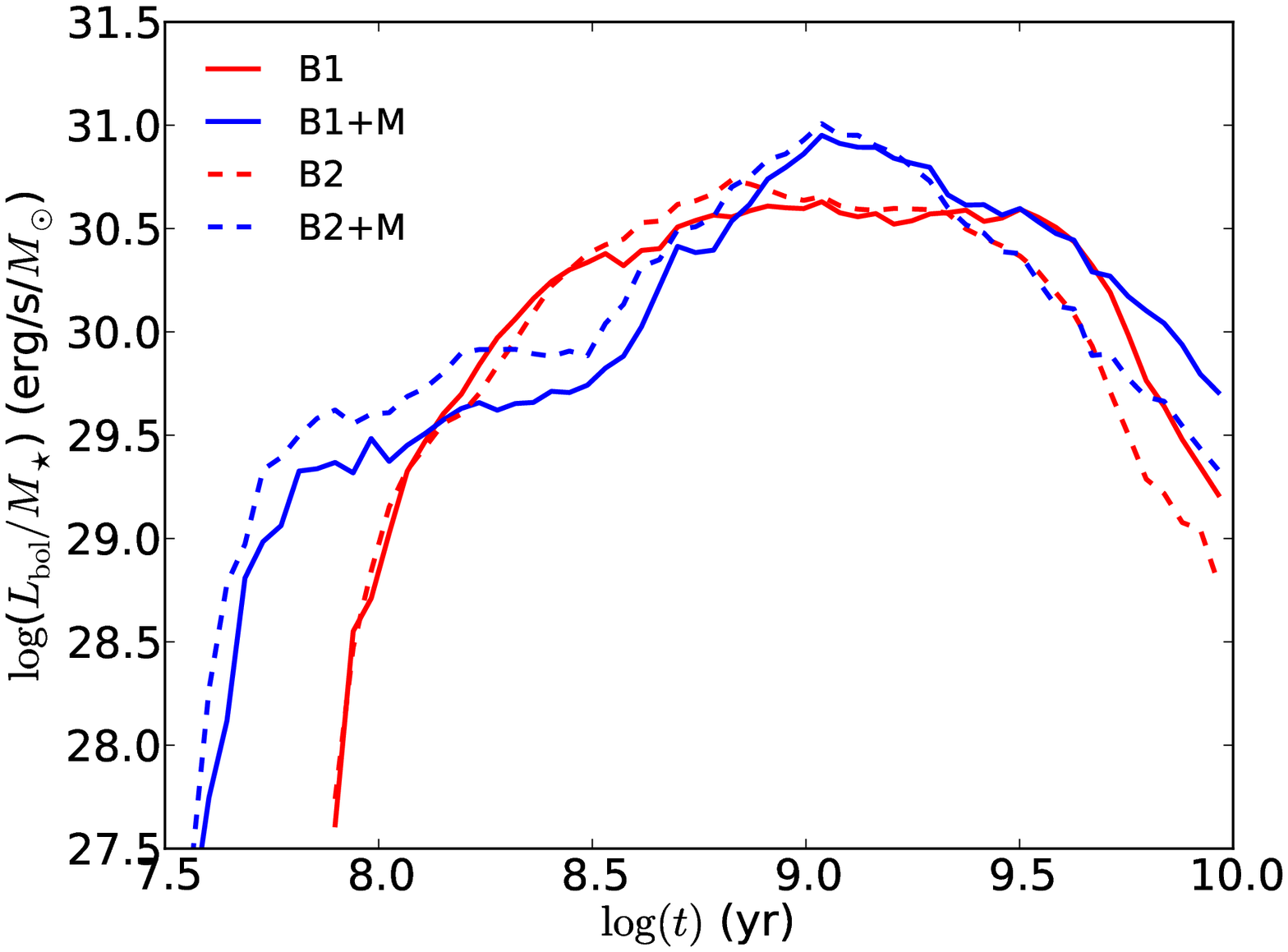}
\includegraphics[width=84mm]{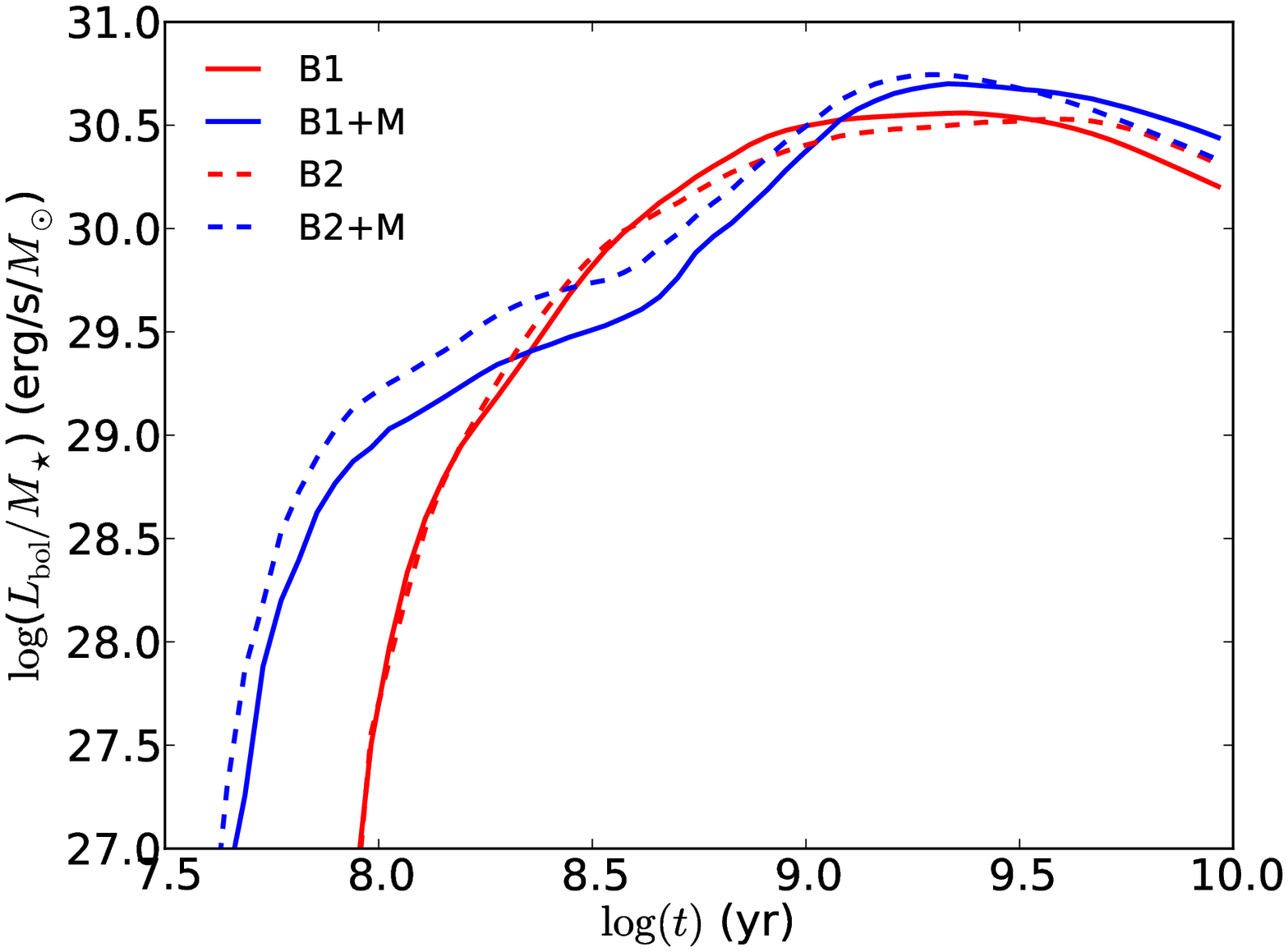}
\caption{ Similar to Fig.~\ref{fig:numsss}, but for bolometric
  luminosity. Upper panel --- starburst case, lower panel --- the case
  of constant SFR (SFR = $1M_{\odot}/$yr). The blue and red lines show the results computed
  with \textsf{BSE+MESA} and \textsf{BSE} only, respectively.  }
\label{fig:bollum}
\end{figure}

\begin{figure}    
\centering
\includegraphics[width=84mm]{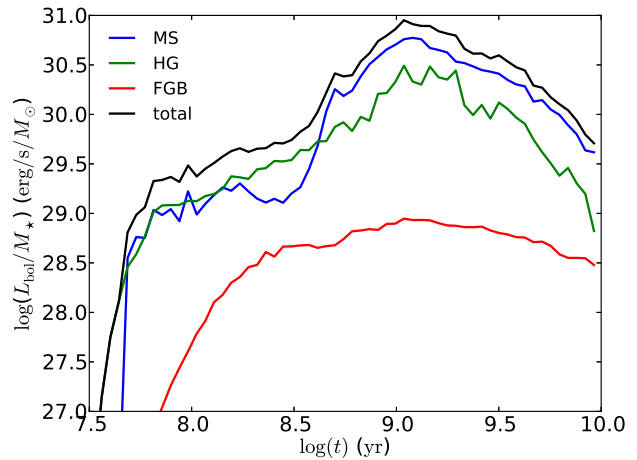}
\caption{Bolometric luminosity of SNBWDs with different types of
  donors for starburst case in model B1+M.}
\label{fig:lumdiffdonor}
\end{figure}

\begin{figure}
\centering
\includegraphics[width=84mm]{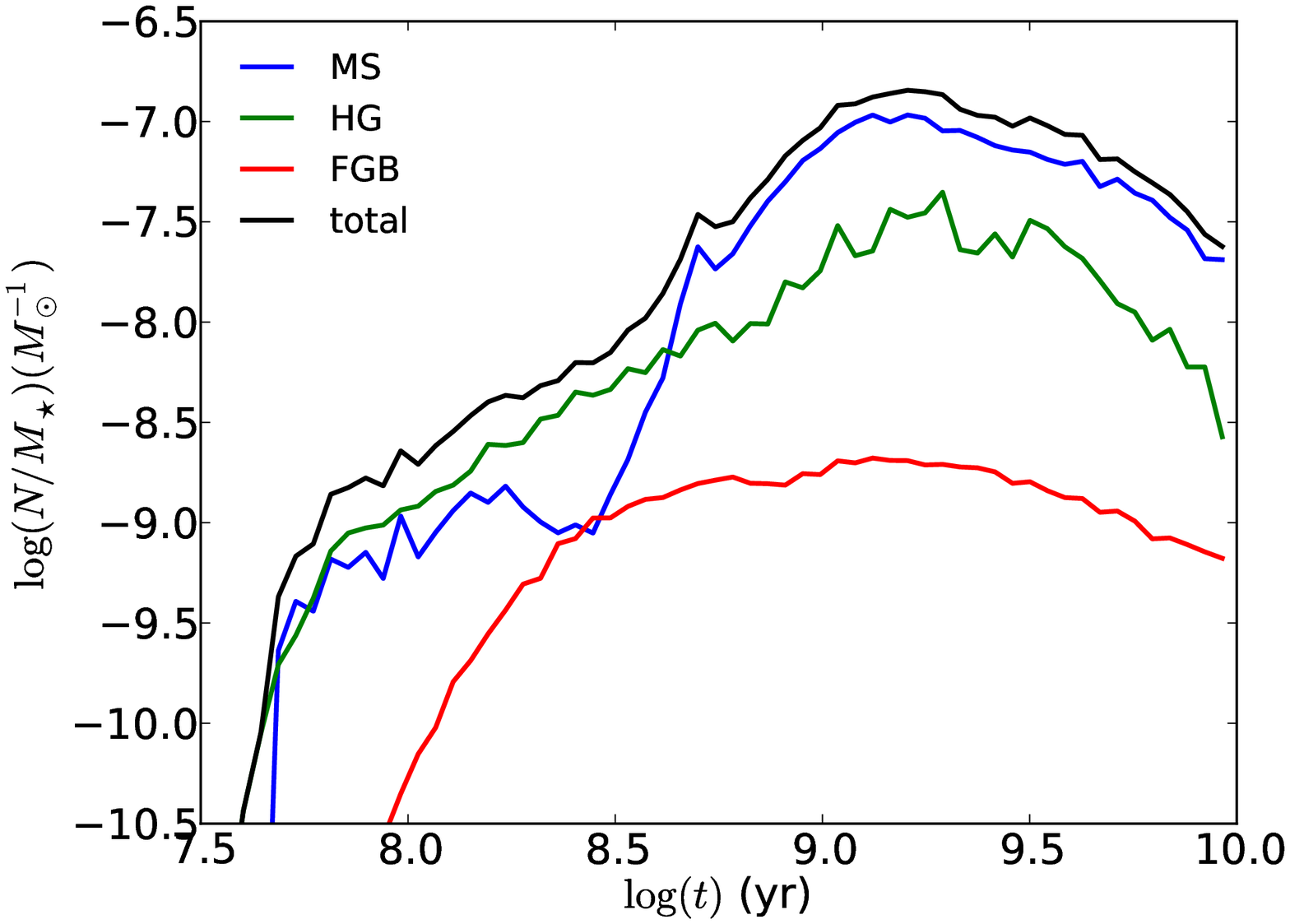}
\caption{Similar to Fig. \ref{fig:lumdiffdonor} but for the number of
  \sws.}
\label{fig:nsssdiffdonor}
\end{figure}

Figures \ref{fig:numsss} and \ref{fig:bollum} show the evolution of
the number of SNBWD per unit mass and their bolometric luminosity, respectively, in
different models. Note immediately, that the difference is not dramatic
between models with precomputed binding energy parameter $\lambda$ and
a fixed \ace\ and models with constant product \al\ in the common
envelope equation. This is unsurprising, given that for relatively low-mass stars $\lambda < 1$,
and does not vary as strongly with evolutionary state as for high-mass stars \citep{xl10,lvk11}. 

As expected from previous studies \citep{y10}, the normalized number of SNBWD in
case I (starburst) is larger than that for case II (constant SFR) at
early age and smaller at late times. In the starburst case, the number of
SNBWDs is sharply decreasing after 2 Gyr, as the reservoir of binaries with ``proper''
combinations of accretor and donor masses is exhausted. In the 
constant SFR case,  SNBWDs form with a delay of about 1\,Gyr respective
to star formation and then ``die'' in about 2\,Gyr, while the galaxy mass continuously increases. This is
the reason for the decrease of $(N/M_\star)$ in the lower panel of Fig.~\ref{fig:numsss}.
At 10\,Gyr, the number of SNBWDs in
\textsf{BSE+MESA} models may be as high as 4550 - 6550 in a
$10^{11} M_{\odot}$ spiral ``galaxy'' and 750 - 1900 in an
elliptical one of the same mass. 
The number of SNBWDs in an elliptical galaxy at $t=10$ Gyr
in our model is comparable to the number in the model of \citet{y10},
while for a spiral galaxy in our model it is a little larger.
Given that most of the soft X-ray emission from SNBWD is easily absorbed by 
interstellar gas, not all SNBWDs will be observed as SSSs. So we must emphasize that
the number of SSS in the model may be estimated only after analysis of the 
spectra of SNBWDs (Paper II, in preparation).

If hydrogen burns steadily on the surface of a WD, its nuclear burning
luminosity is
\begin{equation}
\centering
L_{\rm nuc} = {\epsilon_{\rm H}} {X_{\rm H}} {\dot{M}_{\rm acc}},
\end{equation}
where $\epsilon_{\rm H}=6.4\times10^{18}\mathrm{erg/g}$ is the nuclear energy release per unit mass
of hydrogen, $X_{\rm H}$ is the mass fraction of hydrogen, and $\dot{M}_{\rm acc}$\ is the
accretion rate.

In Figure \ref{fig:bollum}, we present the dependence of the nuclear
luminosity of the \sws\ population on time. As may be
expected, it follows the evolution of the number of \sws.  In
addition, we should note that helium burning contributes
  little to the total bolometric luminosity, since it is only 10\% as
efficient as hydrogen burning.

In Figs.~\ref{fig:lumdiffdonor} and \ref{fig:nsssdiffdonor}, we show the normalized 
bolometric luminosity and numbers of systems
  with donors of different types 
in a starburst ``galaxy''.  We find that at very early times, the
  number of \sws\ and their bolometric luminosity are dominated by HG
  systems, but after about 300\,Myr (approximately the lifetime of a
  3\,\ms\ star on the MS) the population becomes dominated by
  systems with MS-donors. Relative to MS- and HG- donor systems, those binaries which begin mass transfer on the RGB do not play a significant role at any epoch. This is due to the short time they spend in the steady hydrogen
burning regime. However, in describing the number of HG donors, we should note that the donor type is defined at the onset of mass
transfer. In fact, many donors which begin mass transfer on the HG reach the RGB before the end of mass transfer phase. So, this result can not be directly compared with observation. 

\subsubsection{The Number of Rapidly Accreting WDs}
\label{sec:rawd_number}

\begin{figure}
\centering 
\includegraphics[width=84mm]{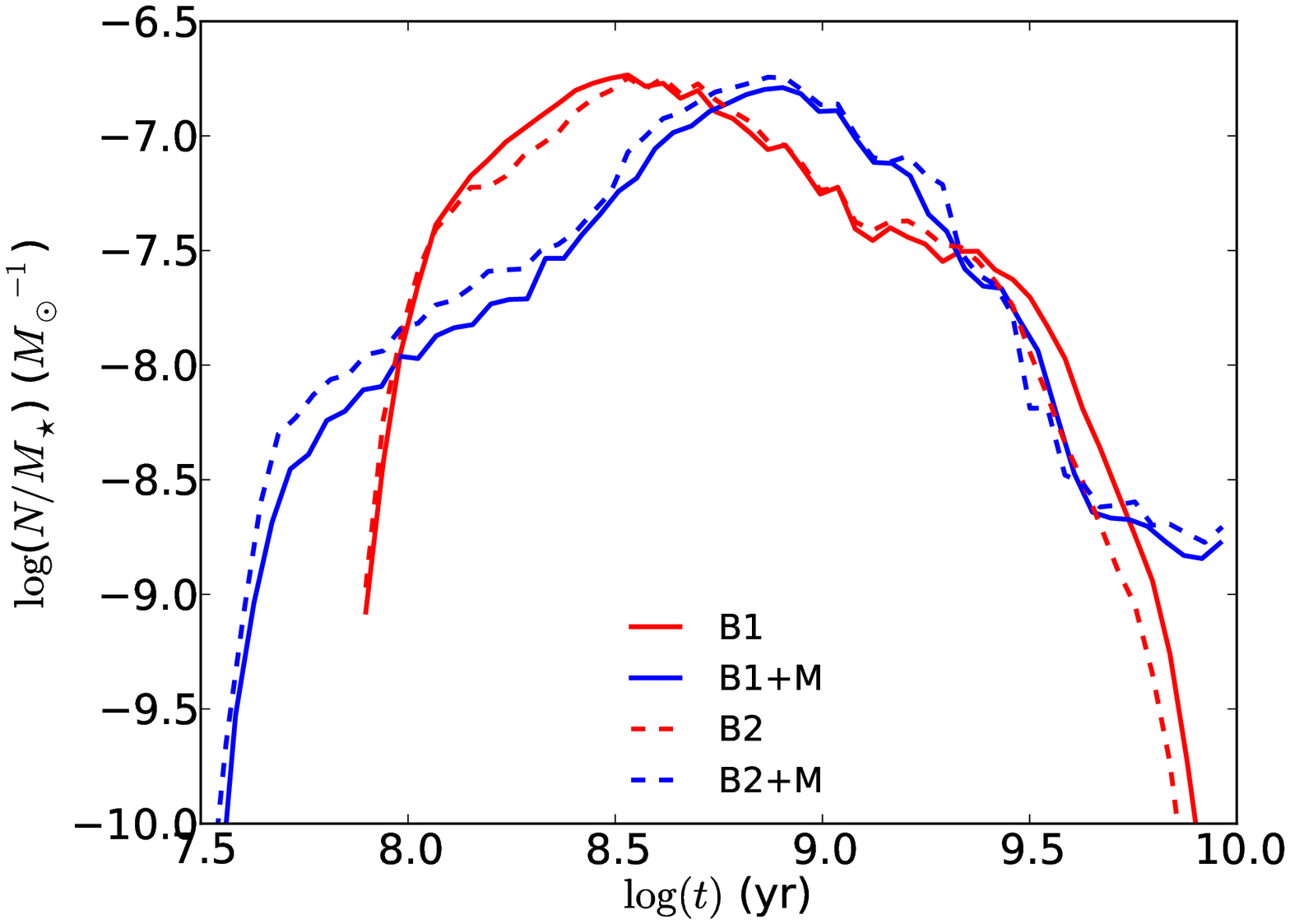}
\includegraphics[width=84mm]{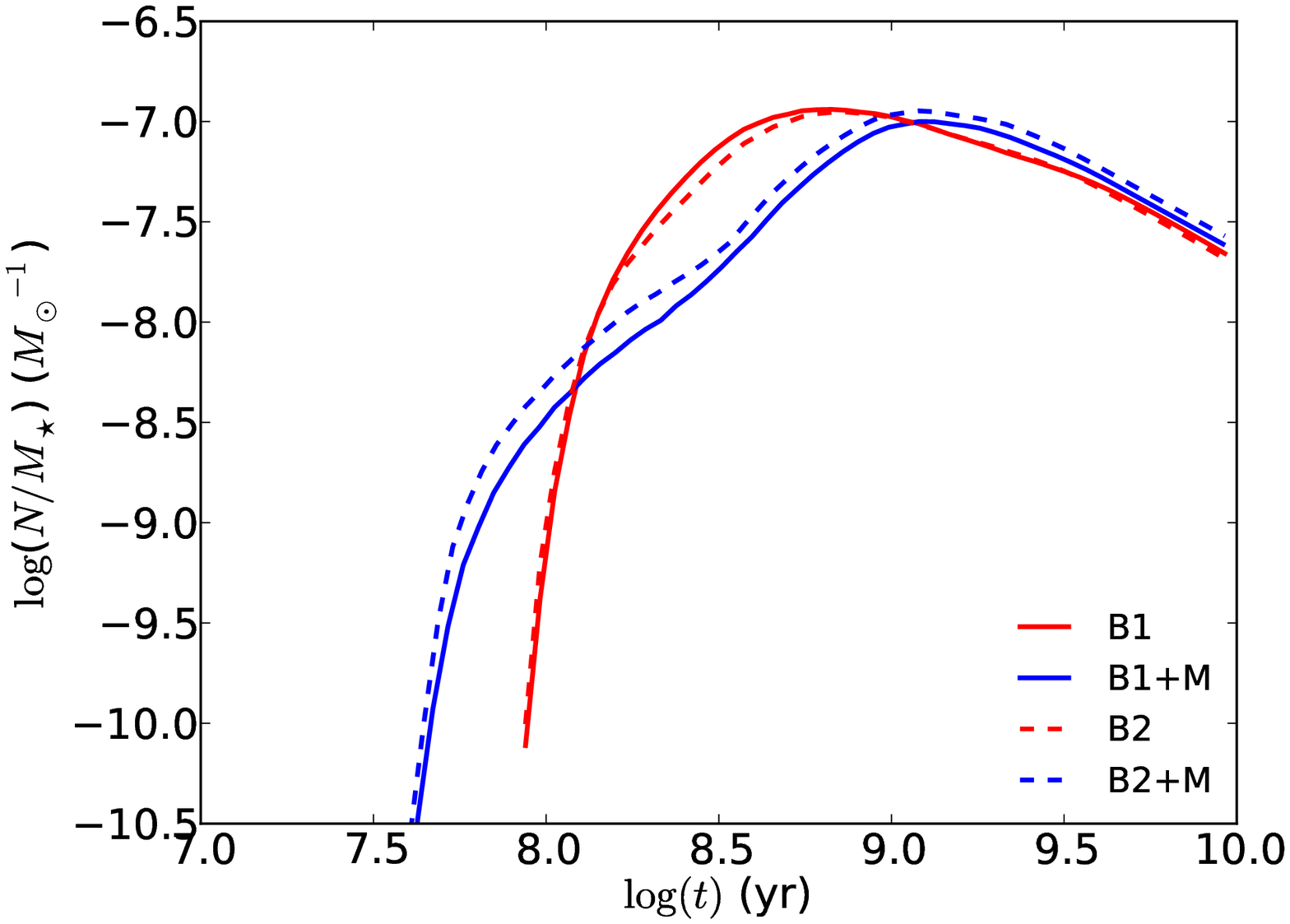}
\caption{The number of RAWDs normalized to the total stellar mass at
  the given time for starburst case (upper panel) and constant SFR
  case with SFR = $1\,M_{\odot}/$yr (lower panel) as a function of time. The blue and red lines
  show the results computed with \textsf{BSE+MESA} and \textsf{BSE}
  only, respectively.}
\label{fig:numrawd}
\end{figure}

The rate of mass transfer is highest at the initial stages of mass
transfer and some NBWD should pass through a RAWD phase
(Fig.~\ref{fig:examples}). In Fig.~\ref{fig:numrawd}, we show the
evolution of the number of RAWDs in each model. It is evident
that the choice in prescription for the binding parameter $\lambda$ (whether fixed or
found from an analytic fit to stellar models) does not lead to the difference in numbers
exceeding $\simeq3$ for RAWDs.

The ``delay'' found between initial star formation and
the formation of RAWDs is shorter in those models which use detailed
calculation of post-RLOF evolution.  This is a consequence of the
difference in the common-envelope formation criterion: in our
BSE-only models systems with massive donors are immediately rejected by the stability criterion, while
in the \textsf{BSE+MESA} model they contribute to the number of RAWDs at very early
epochs. 
The use of the stability criterion from \cite{hw87} in our BSE-only models
 likely results in the lower number of RAWD and NBWD compared to the
\textsf{BSE+MESA}-based calculations. 

The origin of another feature --- the finite time in which RAWDs may exist
  in any starburst ``galaxy'' in models B1, B2 --- may have the
  same reason: while \textsf{BSE} immediately rejects RG-donors on the
  base of the \citet{hw87} criterion for dynamically unstable mass loss,
  the hybrid algorithm always follows the
  increase of \md\ before the formation of a CE, and RAWDs and SSSs should
  inevitably be present in the model, albeit with short lifetimes.

Since the existence of a RAWD phase is based on the assumption that WDs may lose mass through optically thick winds, it
  was suggested that they may be observable as low-luminosity Wolf-Rayet
  stars (possibly WR nuclei of PN?) or V~Sge type cataclysmic binaries with
  numerous emission lines of highly ionized species in their spectra
  \citep[see ][and references therein]{lv13}.  Our ``spiral'' model
  galaxies with mass $10^{11} M_{\odot}$, suggest the existence of 
  2250 - 2500 RAWDs at 10\,Gyr, while in the models of  early-type ``galaxies''
  with the same mass, the number of RAWDs is 160 - 180 at
  10\,Gyr. RAWDs are absent at 10\,Gyr in our BSE-only models, while they remain 
present in our models where the response of the donor is followed throughout the mass transfer phase. 

  This suggests that RAWDs in nearby
    ellipticals, such as M32, can be observed. A search for RAWDs in
  the central core of the Small Magellanic Cloud \citep{lv13} did not
  discover a single RAWD candidate system. Because it is still
  uncertain whether WD may lose mass via optically thick winds, and the
  appearance of RAWD has never been modeled in detail, the nondetection of RAWDs
  does not provide constraints on the SD model of \sne.

   Given that accreting WDs spend a significant time as RAWDs, they can increase 
mass significantly and become progenitors of  SNe~Ia. During the RAWD phase,
the WD photosphere will inflate significantly and emit predominantly
in the extreme UV. Given that this EUV radiation will ionize the surrounding
ISM, \citet{2013MNRAS.432.1640W,2014MNRAS.tmp..350W} 
suggested that observations of emission lines, in particular in He~II~4686\AA, may serve to constrain  the presence of high-temperature ionizing sources.
In order to explore this prediction, \citet{jwgs+14}
selected $\sim 11500$ emission line galaxies and searched for a HeII emission feature.
They found that it is significantly weaker than 
expected if the SD-scenario would be the primary channel for the production of SNe~Ia.
In particular, they found that the contribution of the SD-channel to the total SN~Ia rate
in early-type galaxies with the age $1\mathrm{Gyr}\le t \le 4\mathrm{Gyr}$ must be $<5\%$.

\subsection{SNe Ia rates}
\label{sec:snerates}

\begin{figure}
\centering
\includegraphics[width=84mm]{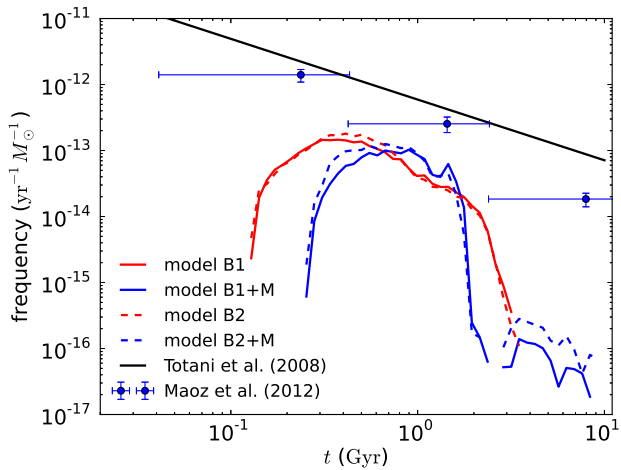}
\caption{Evolution of the SNe Ia rate as a function of galaxy age for elliptical-like galaxy. 
  The power-law line is the fitting formula from \citet{tmod+08} and the points with 
  errorbars are the observed data from \citet{mm12}. }
\label{fig:snburst}
\end{figure}

\begin{figure}
\centering
\includegraphics[width=84mm]{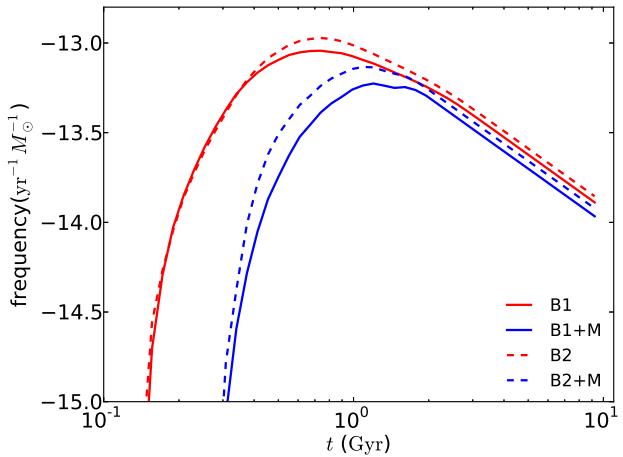}
\caption{Evolution of the SNe Ia rate for spiral-like galaxy with SFR =
  $1.0 M_{\odot}/\mathrm{yr}$.  }
\label{fig:snconst}
\end{figure}

\begin{figure*}
\centering
\includegraphics[width=84mm]{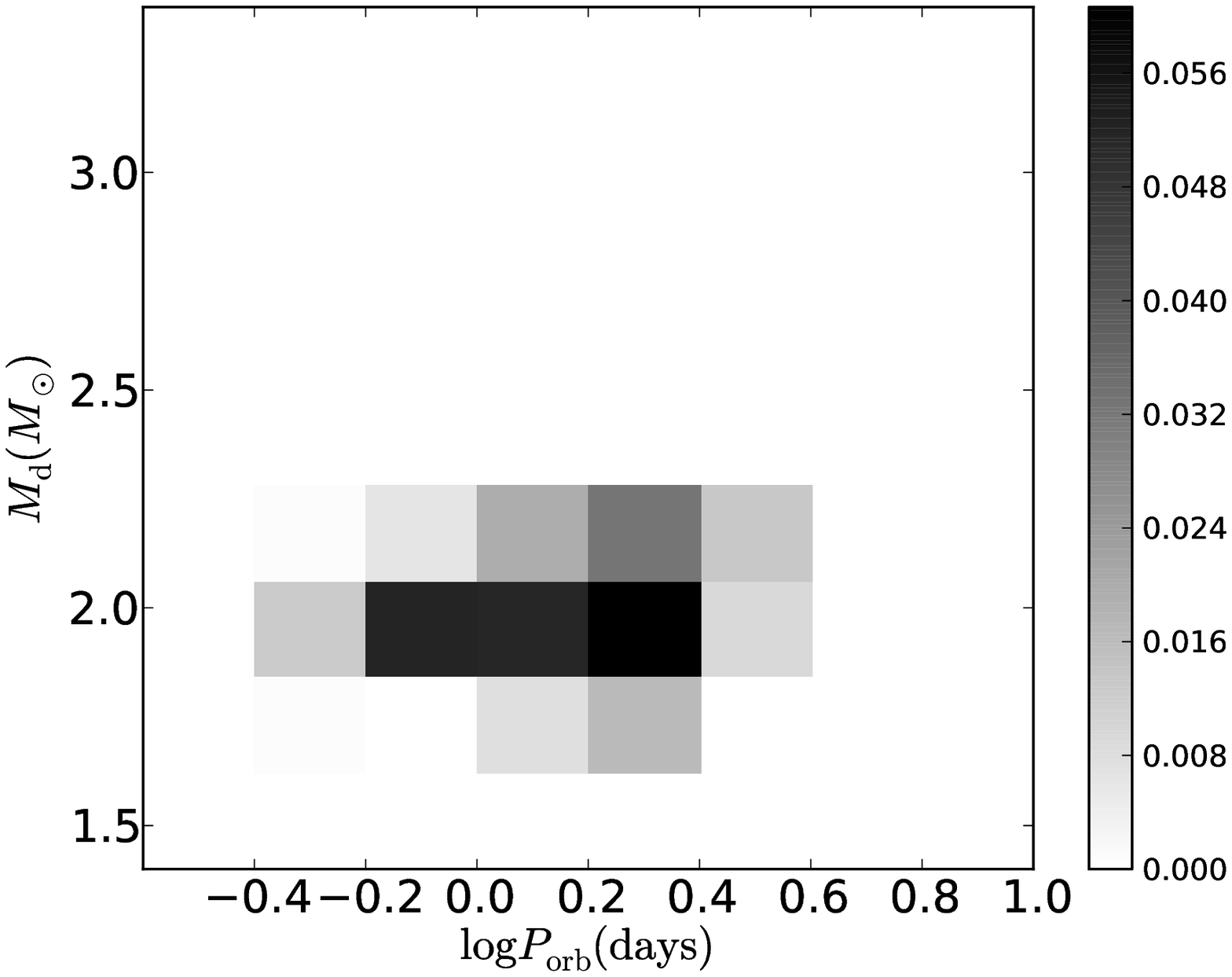}
\includegraphics[width=84mm]{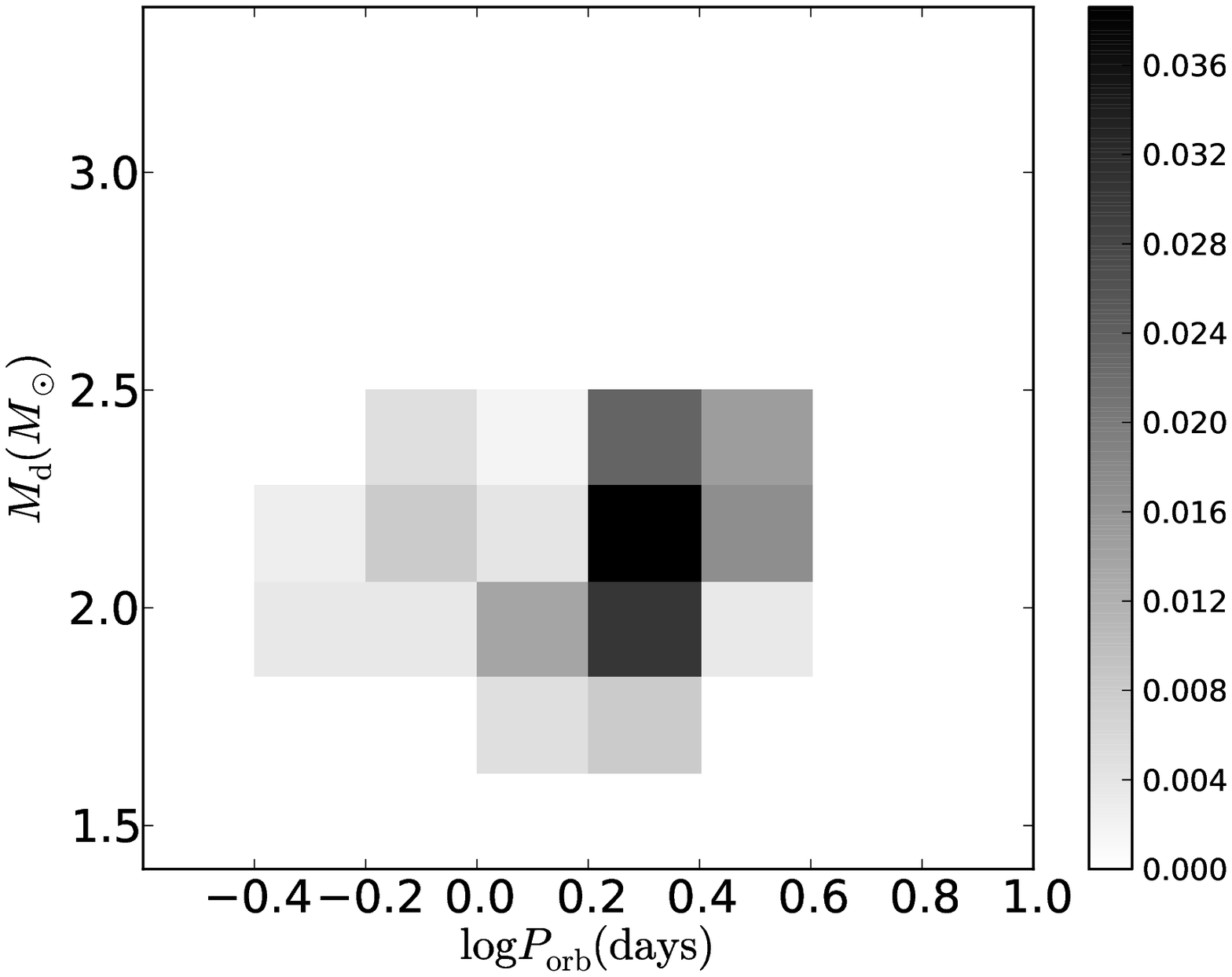}
\includegraphics[width=84mm]{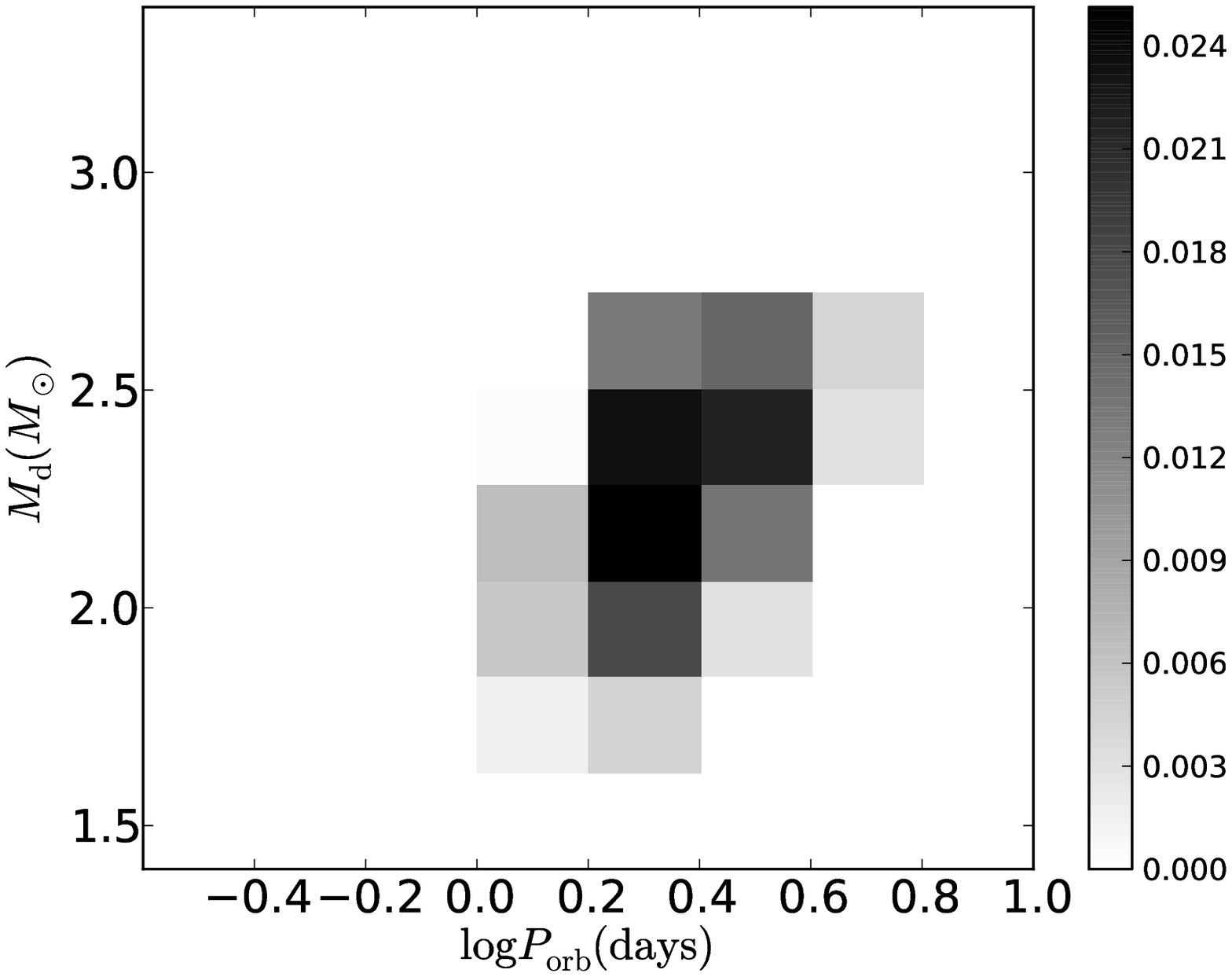}
\includegraphics[width=84mm]{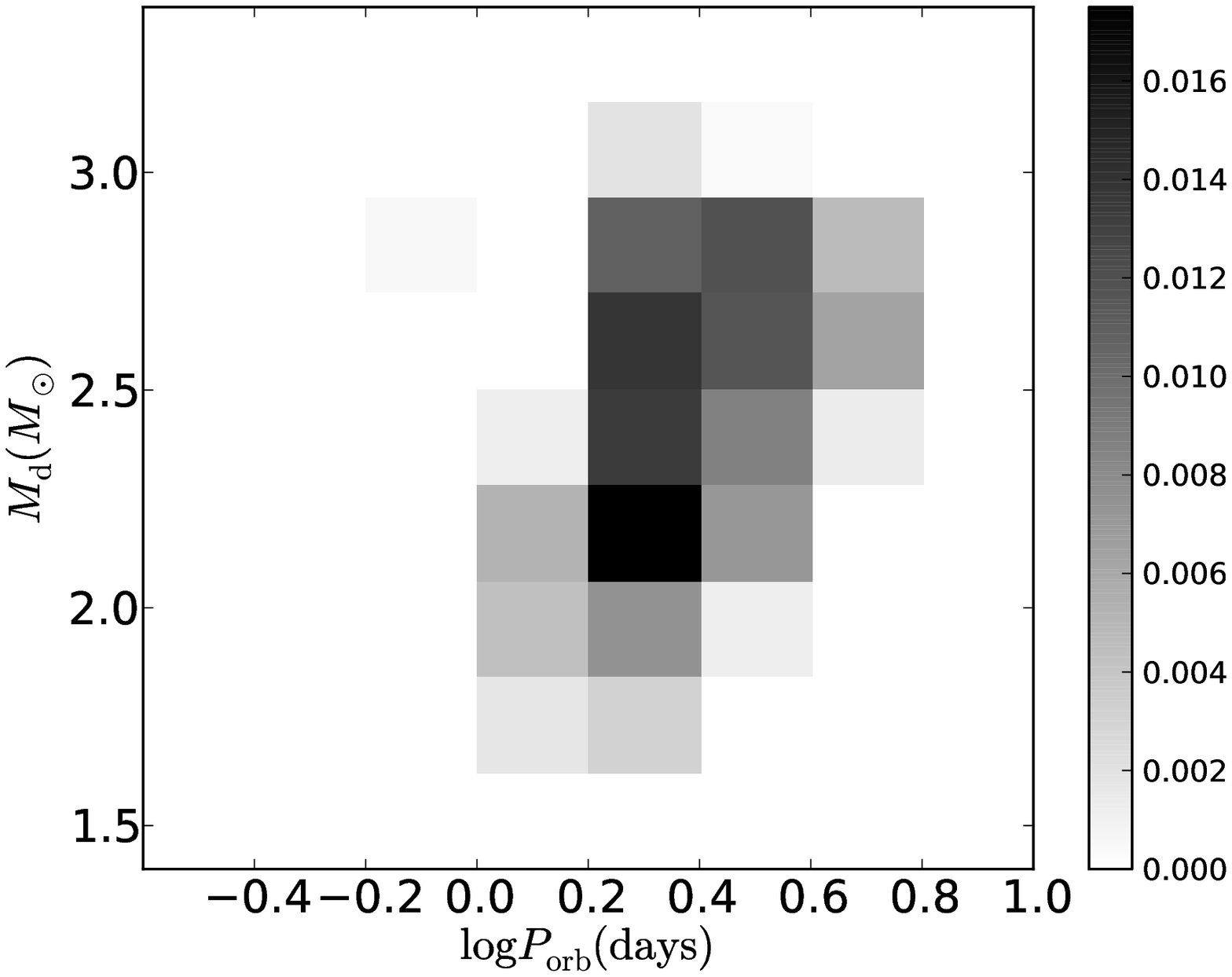}
\caption{The $P_{\mathrm{orb}}-M_{\mathrm{d}}$ distribution at the onset of mass transfer of all successful progenitors of SNe Ia for starbust case 
in model B1+M for different ranges of WD masses: $0.65<= M_{\rm WD} < 0.75$ (upper left panel),
$0.75<= M_{\rm WD} < 0.85$ (upper right), $0.85<= M_{\rm WD} < 0.95$ (lower left) and $0.95<= M_{\rm WD} < 1.05$ (lower right). The gray scale shows the relative contribution of each pixel to the total 
rate of SNe~Ia.
} 
\label{fig:sneinipar}
\end{figure*}

Figure~\ref{fig:snburst} shows the SNe~Ia rate as a function of 
age for a starburst galaxy. Since the delay time distribution (DTD) is
the SN rate as a function of the time elapsed between the formation of a binary
and the explosion of a SN Ia, the plot also presents the DTD.

For comparison, we show the empirical DTDs found by  \cite{tmod+08} and \cite{mmb12}.
From \cite{tmod+08}, we have used their fit
obtained assuming solar abundance and a Salpeter IMF for the stellar population.  
Although we have used different IMF, inspection of their table 4 reveals that the result of \cite{tmod+08} does not strongly 
depend on the IMF. \cite{mmb12} use  Kroupa IMF. Similar to other
studies, the SNe~Ia rate produced by the SD-scenario in our models falls well below the observed one, with
a DTD which does not follow the simple power-law distribution suggested by observations.
It is worth noting that the peak in the DTD for our \textsf{BSE+MESA} models
is shifted to later times, closer to 1\,Gyr, compared to the \textsf{BSE} models.
This is mainly due to the difference in mass-transfer rate treatment, as discussed in the previous subsection.

Our peak calculated SNe~Ia rate for the starburst case 
is still smaller than the observed rate, even though we have adopted 100\% accumulation efficiency for
helium burning. Figure~\ref{fig:snconst} shows
the SNe Ia rate for a spiral-like ``galaxy'' with constant SFR = $1.0
M_{\odot}/\mathrm{yr}$. 
Given that the current and likely for the last many Gyrs SFR in the Milky Way is
$\sim 2 \,\myr$ \citep{chomiuk_sfr11,kennicutt_sfr12}\footnote{ As shown by \citet{chomiuk_sfr11}, modern 
        estimates of current Galactic SFR cluster around $1.9\pm0.4\,\myr$, if normalized to the same assumptions 
        about the IMF of the stellar population
and similar assumptions about stellar evolution are adopted (see Table 1 in the quoted paper).}, 
the SD SNe Ia rate in our simulation is $2.0\times10^{-4} \rm{yr}^{-1}$ at $t=10$ Gyr. This is
$\approx$ 15--20 times lower than the rate inferred for Milky Way like galaxies
$(3-4)\times10^{-3}/\mathrm{yr}$ \citep{cet99}. Note that past estimates of the rate of SNe Ia from the 
    SD-channel \citep[e.g.,][]{hp04} produced larger values partially due to a higher assumed SFR: 3-5\msun/yr.

\citet{hp04,wlh10,my10} adopted a similar method to investigate the SNe Ia rate produced by WD+MS/HG
binaries, but under different assumptions on SFR, binarity fraction, common envelope ejection efficiency, retention 
efficiency and magnetic braking. If we renormalize the rates of SNe Ia found in these papers to a SFR of $2\,\myr$ and binary
fraction 50\%, they do not exceed $3.6\times10^{-4}\rm{yr}^{-1}$ which is not significantly different from the rate which 
we derive. \citet{my10} also investigated the effect of mass stripping and accretion-disk instability, as suggested by
\citet{hkn08} and \citet{krs03}; these hypothetical effects may increase the rate of SNe Ia by a factor of 3-4.

In order to verify the influence of the RAWD phase on the SNe~Ia rate, we also calculated the latter
assuming that the WD binaries will enter CE when mass accretion rate is larger than the maximum rate for stable 
hydrogen burning. We find that SNe~Ia rate becomes negilible in elliptical and spiral galaxies, in accordance with the 
estimates made before the introduction of optically thick winds as a stabilizing effect on mass 
transfer \citep[e.g.,][]{yltt+96}.

In Figure \ref{fig:sneinipar}, we show the distribution of
  binary parameters at the onset of mass transfer for successful progenitors
  of SNe Ia in the starburst case. Here we only show four different
  WD mass ranges, since there are far fewer SNe Ia produced outside of these ranges.
  It is interesting to find that most of the SNe Ia come from binaries with
  initially less massive WDs, which is consistent with \cite{mch09} (see their Fig. 9). 
This is not difficult to understand, since most white dwarf binaries 
in the post-CE white dwarf binary population have relatively 
  small-mass WDs \citep{hpe95}.
  Moreover, the lower mass WDs can accrete more
  mass in binaries with the same donor mass and orbital periods \citep{ldwh00}.
Note that contribution of systems with red giant donors is insignificant in our 
calculation. Note, in our calculations, we do not consider possible atmospheric mass loss
\citep{pup73} or enhancement of stellar winds of RG close to ROLF \citep{pm07}, which may
delay embedding of the potentially unstable system into common envelope at the instant of
RLOF \citep[e.g.][]{pm07}.

\subsection{Uncertainty of common envelope evolution}
\begin{figure}
\centering
\includegraphics[width=90mm]{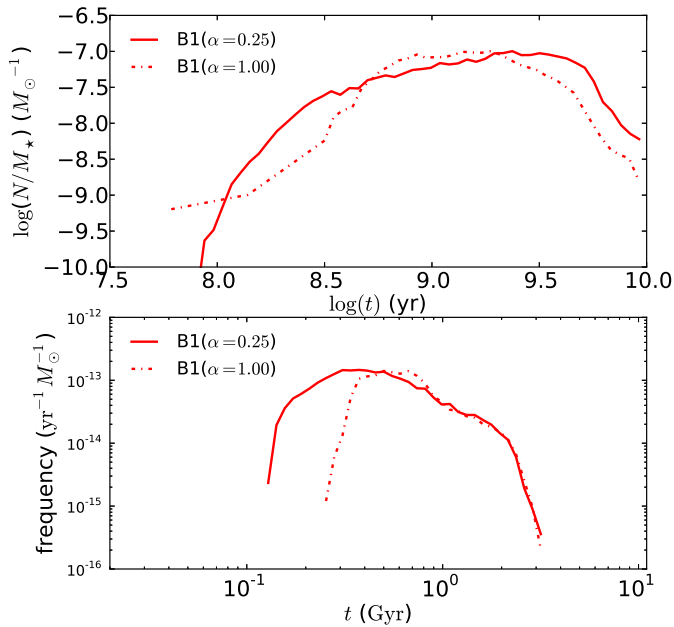}
\caption{Comparision of mass-normalized SNBWDs number (upper panel) and SNe Ia rate (lower panel) for starburst case
    in the default configuration (solid line) and assuming $\alpha=1.0$ (dash-dotted line).
}
\label{fig:num_sss_sne_diff_alpha}
\end{figure}

It is widely understood that the outcome of CE evolution suffers from many uncertainties, such as the available sources of energy \citep[e.g.][]{ivanova_al_ce13,zsp14}. In our calculation, we adopt $\alpha = 0.25$ and a fitting formula
for $\lambda$. In the fitting formula for $\lambda$, the internal energy such as thermal energy of the gas and 
radiation energy are included \citep[see Eq. (1) in][]{lvk11}. Given the uncertainty of CE 
evolution, we performed a set of calculations of the model B1 increasing $\alpha$ to 1. 
In Fig. \ref{fig:num_sss_sne_diff_alpha}, we compare the evolution of SNBWDs
and SNe Ia rate in the starburst case for $\alpha = 0.25$ and $\alpha = 1.0$. We found that there is no dramatic
difference between $\alpha= 1.0$ and $\alpha = 0.25$. For  $\alpha = 1.0$, 
SNBWD, as well as SN~Ia events appear slightly later than for  $\alpha = 0.25$,
since after the first common envelope episode systems are wider. A
similar effect was noted before by \citet{wlh10}. We also found no significant difference for RAWDs, which 
are not shown here. 

\subsection{Remarks about the noise in population synthesis calculations}

In the above plots, especially Fig.~\ref{fig:lumdiffdonor}, ~\ref{fig:nsssdiffdonor}   and~\ref{fig:snburst},
one may note that the curves are noisy. This problem is commonly seen in population synthesis studies.
There are several reasons for that, both in our \textsf{BSE}-only calculation as well as in our \textsf{BSE}+\textsf{MESA}
calculation, the primary causes being:

\begin{itemize}

\item In our \textsf{BSE}-only calculation, one cause for the noise is that we have adopted several different CE criteria
for different binaries. This is especially important for those binaries that begin mass transfer on the HG, which evolve to the RG phase prior to the end of the mass transfer phase. 
Typically, our HG-donor tracks end abruptly when they violate the stability condition of \cite{hw87}. However, just prior to this, many tracks ``dip back'' into the stable burning regime. Whether this occurs or not does not depend smoothly on the time of onset of mass transfer, therefore introducing small variations in the numbers and luminosity of SNBWDs predicted in our calculations. 

\item In addition to this, the limited number of tracks in the calculation will also contribute to the noise. In particular, in our \textsf{BSE}+\textsf{MESA} calculations, we perform a mapping from a set of $\sim$ 100,000 tracks produced by \textsf{BSE}, to one of $\sim$ 30,000 tracks produced by MESA. The discontinuous transition from our \textsf{BSE} grid to a much coarse grid of \textsf{MESA} tracks introduces unphysical variability in our output.

\item Finally, the noise in the Fig. \ref{fig:snburst} is primarily due to the small number  of binary tracks which explode as SNe Ia.

\end{itemize}

\section{Summary and Conclusions}
\label{sec:sum}

In this work, we combined the population synthesis code \textsf{BSE} with the detailed stellar evolutionary
code \textsf{MESA} for the first time, in order to study the population of accreting WDs. We also compared the output from this
with the results obtained applying a ``rapid'' algorithm, using \textsf{BSE} alone. With these two BPS algorithms, we
investigated the evolution of the number of rapidly accreting white dwarfs, stable nuclear burning white dwarfs and the SNe Ia rate in elliptical and spiral-like galaxies. In addition to confirming that the SD channel is subdominant in producing the overall SN Ia rate, we also evaluated the effect of implementing differing treatments of mass transfer for the results of BPS calculations.

Comparing the two versions of our binary population synthesis calculations, we found that the
mass transfer prescription in BPS is especially important for calculating the number and total luminosity of nuclear-burning white dwarfs in elliptical galaxies at early and late epochs (from initial starburst). We argue that this also partially explains the differences
in SNe~Ia rates and DTD obtained by different binary population synthesis groups.
We found that RAWDs appear earlier in our \textsf{BSE+MESA} model compared with
\textsf{BSE}-only, due to the accreting WDs with massive donors found in the former. 

We find that there is a factor of $\approx$ 3 difference between the results of our calculations 
using a fitting formula for the binding parameter $\lambda$ \citep{lvk11} and a constant $\alpha_{\rm ce}=0.25$,  and our calculations using a constant $\alpha_{\rm ce} \times \lambda=0.25$.

In our \textsf{BSE+MESA} model, we found that the number of RAWDs at 10
Gyr is $160 - 180$ for an elliptical galaxy of $10^{11}$\,\ms\ and $2250 - 2500$ for a
spiral-like galaxy of the same mass. 
This result  is in stark contrast with zero RAWDs predicted in our calculation 
for a model elliptical galaxy using \textsf{BSE} alone.
 
We find that the number of SNBWD is $750 - 1900$ in our model 
elliptical-like galaxy and $4550 - 6550$ in our model spiral-like galaxy (both with $M=10^{11}$\,\ms\,at 10 Gyr).

The predicted SD SNe Ia rate for a Milky-Way-like galaxy is found to be $\simeq2.0\times 10^{-4} \mathrm{yr}^{-1}$, more than an
order of magnitude lower than the observationally inferred 
total Galactic SNe~Ia rate. Our DTD for the SD-model is inconsistent with the observed DTD $\propto t^{-1}$. If we assume that RAWDs do not exist, but rather that a common envelope
is formed if the accretion rate onto a WD is larger that the upper limit for stable hydrogen burning, then the rate of \sne\ produced by the SD-channel becomes negligible.

To conclude our discussion of the significance of the SD-channel for the production of \sne, we note the following. 
Since WDs in the steady burning phase can effectively accumulate
mass, it is widely suggested that SSSs be
identified with the progenitors of SNe Ia and,
therefore, observations of SSSs may be useful for constraining the SD model.
\citet{gb10} estimated the expected X-ray flux from the progenitors of SNe Ia in the SD-scenario 
for elliptical galaxies based on the observed supernova rate. They found that the observed X-ray flux from six
nearby elliptical galaxies is $30-50$ times smaller than the predicted value,
and constrain the contribution of the SD-channel to $<5\%$. In a similar way, 
\citet{ste10a} found that there are too few SSSs to account for the SNe Ia rate. Even if accreting WDs radiate at significantly
lower temperatures (T$\approx 10^5K$), due to the inflation of their photospheres, the SD channel may still be limited to providing
$< 10\%$ of the \sne\ rate \citep{2013MNRAS.432.1640W,jwgs+14}.

\citet{yltt+96}, \citet{ste10a}, and \citet{y10} considered the possibility that some SSSs may reside in wind-accreting 
systems which later produce  double-degenerate systems (symbiotic stars), and estimated their number 
in the Galaxy as  $\sim 10^3$, while the estimate by \citet{nnvt14} is even lower:  $\sim 10^2$. 
Though several $10^3$\ SNBWDs in the model is apparently a large number, it is still about  2 orders of magnitude lower than that necessary 
to be consistent with the
observationally inferred  Galactic \sne\ rate.  
\citet{2006MNRAS.372.1389L} and \citet{nnvt14} estimated that in wind-fed systems WD typically accrete 
no more than $\approx$0.1\ms.   

\section*{Acknowledgments}
We would like to thank an anonymous referee for useful comments,
which helped to improve the paper.
HLC gratefully acknowledges support and hospitality from the
MPG-CAS Joint Doctoral Promotion Program (DPP) and Max Planck
Institute for Astrophysics (MPA). This work of HLC and ZWH
was partially supported by the National Natural Science Foundation
of China (Grant Nos. 11033008,11390374). The work is also partially supported by
Presidium of the Russian Academy of Sciences P-21 and RFBR grant 
14-02-00604. LRY gratefully acknowledges warm hospitality and support from MPA-Garching. 
MG acknowledges hospitality of the Kazan Federal University (KFU) and support by the Russian Government Program of Competitive Growth of KFU. HLC acknowledges the computing time granted by the Yunnan Observatories and 
provided on the facilities at the Yunnan Observatories Supercomputing Platform.

\def\apj{ApJ} \def\apjl{ApJL} \def\apjs{ApJS} \def\aj{AJ}
\def\aap{A\&A} \def\araa{ARA\&A} \def\aapss{A\&AS} \def\mnras{MNRAS}
\def\nat{Nature} \def\nar{New A Rev.}  \def\apss{Ap\&SS}
\def\pasp{PASP} \def\pasa{PASA} \def\pasj{PASJ} 
\def\acta{Acta. Astron.}
\def\memsai{Mem. Soc. Astron. Italiana}
\def\aapr{Astron. Astrophys. Rev.}
\def\sovast{SvA}
\def\actaa{Acta. Astron.}

\bibliographystyle{mn2e} 

\begin{thebibliography}{74}
\expandafter\ifx\csname natexlab\endcsname\relax\def\natexlab#1{#1}\fi

\bibitem[{{Abt}(1983)}]{abt83}
{Abt} H.~A., 1983, \araa, 21, 343

\bibitem[{{Bloom} {et~al}\mbox{.}(2012){Bloom}, {Kasen}, {Shen}, {Nugent},
  {Butler}, {Graham}, {Howell}, {Kolb}, {Holmes}, {Haswell}, {Burwitz},
  {Rodriguez}, \& {Sullivan}}]{bksb+12}
{Bloom} J.~S. {et~al.}, 2012, \apjl, 744, L17

\bibitem[{{Bours}, {Toonen} \& {Nelemans}(2013){Bours}, {Toonen}, \&
  {Nelemans}}]{btn13}
{Bours} M.~C.~P., {Toonen} S., {Nelemans} G., 2013, \aap, 552, A24

\bibitem[{{Cappellaro}, {Evans} \& {Turatto}(1999){Cappellaro}, {Evans}, \&
  {Turatto}}]{cet99}
{Cappellaro} E., {Evans} R., {Turatto} M., 1999, \aap, 351, 459

\bibitem[{{Chomiuk} \& {Povich}(2011)}]{chomiuk_sfr11}
{Chomiuk} L., {Povich} M.~S., 2011, \aj, 142, 197

\bibitem[{{Davis}, {Kolb} \& {Willems}(2010){Davis}, {Kolb}, \&
  {Willems}}]{dkw10}
{Davis} P.~J., {Kolb} U., {Willems} B., 2010, \mnras, 403, 179

\bibitem[{de~Kool(1990)}]{dek90}
de~Kool M., 1990, {\apj}, 358, 189

\bibitem[{{Dewi} \& {Tauris}(2000)}]{dt00}
{Dewi} J.~D.~M., {Tauris} T.~M., 2000, \aap, 360, 1043

\bibitem[Di Stefano(2010)]{ste10a} 
Di Stefano, R.\ 2010, \apj, 719, 474 

\bibitem[{Duquennoy \& Mayor(1991)}]{Duquennoy_Mayor91}
Duquennoy A., Mayor M., 1991, Astron. Astrophys., 248, 485

\bibitem[{{Gilfanov} \& {Bogd{\'a}n}(2010)}]{gb10}
{Gilfanov} M., {Bogd{\'a}n} {\'A}., 2010, \nat, 463, 924

\bibitem[{{Hachisu}, {Kato} \& {Nomoto}(1999){Hachisu}, {Kato}, \&
  {Nomoto}}]{hkn99}
{Hachisu} I., {Kato} M., {Nomoto} K., 1999, \apj, 522, 487

\bibitem[Hachisu et al.(2008)]{hkn08} Hachisu, I., Kato, M., 
\& Nomoto, K.\ 2008, \apj, 679, 1390 

\bibitem[{{Han} \& {Podsiadlowski}(2004)}]{hp04}
{Han} Z., {Podsiadlowski} P., 2004, \mnras, 350, 1301

\bibitem[{{Han}, {Podsiadlowski} \& {Eggleton}(1995){Han}, {Podsiadlowski}, \&
  {Eggleton}}]{hpe95}
{Han} Z., {Podsiadlowski} P., {Eggleton} P.~P., 1995, \mnras, 272, 800

\bibitem[{{Hillebrandt} {et~al}\mbox{.}(2013){Hillebrandt}, {Kromer},
  {R{\"o}pke}, \& {Ruiter}}]{hkrr13}
{Hillebrandt} W., {Kromer} M., {R{\"o}pke} F.~K., {Ruiter} A.~J., 2013,
  Frontiers of Physics, 8, 116

\bibitem[{{Hjellming} \& {Webbink}(1987)}]{hw87}
{Hjellming} M.~S., {Webbink} R.~F., 1987, \apj, 318, 794

\bibitem[{{Hurley}, {Tout} \& {Pols}(2002){Hurley}, {Tout}, \& {Pols}}]{htp02}
{Hurley} J.~R., {Tout} C.~A., {Pols} O.~R., 2002, \mnras, 329, 897
 their 
\bibitem[{{Iben} \& {Tutukov}(1984)}]{it84}
{Iben}, Jr. I., {Tutukov} A.~V., 1984, \apjs, 54, 335

\bibitem[{{Iben} \& {Tutukov}(1989)}]{it89}
{Iben}, Jr. I., {Tutukov} A.~V., 1989, \apj, 342, 430

\bibitem[{{Idan}, {Shaviv} \& {Shaviv}(2013){Idan}, {Shaviv}, \&
  {Shaviv}}]{iss13}
{Idan} I., {Shaviv} N.~J., {Shaviv} G., 2013, \mnras, 433, 2884

\bibitem[{{Ivanova} {et~al}\mbox{.}(2013){Ivanova}, {Justham}, {Chen}, {De
  Marco}, {Fryer}, {Gaburov}, {Ge}, {Glebbeek}, {Han}, {Li}, {Lu}, {Marsh},
  {Podsiadlowski}, {Potter}, {Soker}, {Taam}, {Tauris}, {van den Heuvel}, \&
  {Webbink}}]{ivanova_al_ce13}
{Ivanova} N. {et~al.}, 2013, \aapr, 21, 59

\bibitem[{{Ivanova} \& {Taam}(2004)}]{Ivanova04}
{Ivanova} N., {Taam} R.~E., 2004, \apj, 601, 1058

\bibitem[{{Johansson} {et~al}\mbox{.}(2014){Johansson}, {Woods}, {Gilfanov},
  {Sarzi}, {Chen}, \& {Oh}}]{jwgs+14}
{Johansson} J., {Woods} T.~E., {Gilfanov} M., {Sarzi} M., {Chen} Y.-M., {Oh}
  K., 2014, ArXiv e-prints

\bibitem[{{Kennicutt} \& {Evans}(2012)}]{kennicutt_sfr12}
{Kennicutt} R.~C., {Evans} N.~J., 2012, \araa, 50, 531

\bibitem[{{Kraicheva} {et~al}\mbox{.}(1979){Kraicheva}, {Popova}, {Tutukov}, \&
  {Yungelson}}]{1979SvA....23..290K}
{Kraicheva} Z.~T., {Popova} E.~I., {Tutukov} A.~V., {Yungelson} L.~R., 1979,
  \sovast, 23, 290
 
\bibitem[King et al.(2003)]{krs03} King, A.~R., Rolfe, D.~J., 
  \& Schenker, K.\ 2003, \mnras, 341, L35 

\bibitem[{{Kroupa}(2001)}]{krou01}
{Kroupa} P., 2001, \mnras, 322, 231

\bibitem[{{Langer} {et~al}\mbox{.}(2000){Langer}, {Deutschmann}, {Wellstein},
  \& {H{\"o}flich}}]{ldwh00}
{Langer} N., {Deutschmann} A., {Wellstein} S., {H{\"o}flich} P., 2000, \aap,
  362, 1046

\bibitem[{{Lepo} \& {van Kerkwijk}(2013)}]{lv13}
{Lepo} K., {van Kerkwijk} M., 2013, \apj, 771, 13

\bibitem[{{Lin} {et~al}\mbox{.}(2011){Lin}, {Rappaport}, {Podsiadlowski},
  {Nelson}, {Paxton}, \& {Todorov}}]{2011ApJ...732...70L}
{Lin} J., {Rappaport} S., {Podsiadlowski} P., {Nelson} L., {Paxton} B.,
  {Todorov} P., 2011, \apj, 732, 70

\bibitem[{{Loveridge}, {van der Sluys} \& {Kalogera}(2011){Loveridge}, {van der
  Sluys}, \& {Kalogera}}]{lvk11}
{Loveridge} A.~J., {van der Sluys} M.~V., {Kalogera} V., 2011, \apj, 743, 49

\bibitem[L{\"u} et al.(2006)]{2006MNRAS.372.1389L} L{\"u}, G., Yungelson, 
L., \& Han, Z.\ 2006, \mnras, 372, 1389 

\bibitem[Madhusudhan et al.(2008)]{mrpn08} Madhusudhan, N., 
Rappaport, S., Podsiadlowski, P., \& Nelson, L.\ 2008, \apj, 688, 1235 

\bibitem[{{Maoz} \& {Mannucci}(2012)}]{mm12}
{Maoz} D., {Mannucci} F., 2012, \pasa, 29, 447

\bibitem[{{Maoz}, {Mannucci} \& {Brandt}(2012){Maoz}, {Mannucci}, \&
  {Brandt}}]{mmb12}
{Maoz} D., {Mannucci} F., {Brandt} T.~D., 2012, \mnras, 426, 3282

\bibitem[{{Maoz}, {Mannucci} \& {Nelemans}(2013){Maoz}, {Mannucci}, \&
  {Nelemans}}]{mmn13}
{Maoz} D., {Mannucci} F., {Nelemans} G., 2013, ArXiv e-prints

\bibitem[{{Matteucci} \& {Greggio}(1986)}]{mg86}
{Matteucci} F., {Greggio} L., 1986, \aap, 154, 279

\bibitem[Meng \& Yang(2010)]{my10} 
    Meng, X., \& Yang, W.\ 2010, \apj, 710, 1310 

\bibitem[{{Meng}, {Chen} \& {Han}(2009){Meng}, {Chen}, \& {Han}}]{mch09}
{Meng} X., {Chen} X., {Han} Z., 2009, \mnras, 395, 2103

\bibitem[{{Morton}(1960)}]{1960ApJ...132..146M}
{Morton} D.~C., 1960, \apj, 132, 146

\bibitem[{{Nelson}(2012)}]{2012JPhCS.341a2008N}
{Nelson} L., 2012, Journal of Physics Conference Series, 341, 012008

\bibitem[Newsham et al.(2013)]{nst13} Newsham, G., 
Starrfield, S., \& Timmes, F.\ 2013, arXiv:1303.3642 

\bibitem[Nielsen et al.(2014)]{nnvt14} 
Nielsen, M.~T.~B., Nelemans, G., Voss, R., \& Toonen, S.\ 2014, \aap, 563, A16 

\bibitem[{{Nugent} {et~al}\mbox{.}(2011){Nugent}, {Sullivan}, {Cenko},
  {Thomas}, {Kasen}, {Howell}, {Bersier}, {Bloom}, {Kulkarni}, {Kandrashoff},
  {Filippenko}, {Silverman}, {Marcy}, {Howard}, {Isaacson}, {Maguire},
  {Suzuki}, {Tarlton}, {Pan}, {Bildsten}, {Fulton}, {Parrent}, {Sand},
  {Podsiadlowski}, {Bianco}, {Dilday}, {Graham}, {Lyman}, {James}, {Kasliwal},
  {Law}, {Quimby}, {Hook}, {Walker}, {Mazzali}, {Pian}, {Ofek}, {Gal-Yam}, \&
  {Poznanski}}]{nsct+11}
{Nugent} P.~E. {et~al.}, 2011, \nat, 480, 344

\bibitem[Paczy{\'n}ski(1971)]{1971AcA....21..417P} Paczy{\'n}ski, B.\ 1971, 
\actaa, 21, 417 

\bibitem[{{Paczy{\' n}ski}, {Zi{\'o}lkowski} \& {Zytkow}(1969){Paczy{\' n}ski},
  {Zi{\'o}lkowski}, \& {Zytkow}}]{pzz69}
{Paczy{\' n}ski} B., {Zi{\'o}lkowski} J., {Zytkow} A., 1969, in Mass Loss from
  Stars, p. 237

\bibitem[{Paczy\'nski \& Sienkiewicz(1972)}]{ps72a}
Paczy\'nski B., Sienkiewicz R., 1972, \acta, 21, 1

\bibitem[{{Passy}, {Herwig} \& {Paxton}(2012){Passy}, {Herwig}, \&
  {Paxton}}]{php12}
{Passy} J.-C., {Herwig} F., {Paxton} B., 2012, \apj, 760, 90

\bibitem[{{Paxton} {et~al}\mbox{.}(2011){Paxton}, {Bildsten}, {Dotter},
  {Herwig}, {Lesaffre}, \& {Timmes}}]{pbdh+11}
{Paxton} B., {Bildsten} L., {Dotter} A., {Herwig} F., {Lesaffre} P., {Timmes}
  F., 2011, \apjs, 192, 3

\bibitem[{{Paxton} {et~al}\mbox{.}(2013){Paxton}, {Cantiello}, {Arras},
  {Bildsten}, {Brown}, {Dotter}, {Mankovich}, {Montgomery}, {Stello}, {Timmes},
  \& {Townsend}}]{pcab+13}
{Paxton} B. {et~al.}, 2013, \apjs, 208, 4

\bibitem[{{Perlmutter} {et~al}\mbox{.}(1999){Perlmutter}, {Aldering},
  {Goldhaber}, {Knop}, {Nugent}, {Castro}, {Deustua}, {Fabbro}, {Goobar},
  {Groom}, {Hook}, {Kim}, {Kim}, {Lee}, {Nunes}, {Pain}, {Pennypacker},
  {Quimby}, {Lidman}, {Ellis}, {Irwin}, {McMahon}, {Ruiz-Lapuente}, {Walton},
  {Schaefer}, {Boyle}, {Filippenko}, {Matheson}, {Fruchter}, {Panagia},
  {Newberg}, {Couch}, \& {Supernova Cosmology Project}}]{pagk+99}
{Perlmutter} S. {et~al.}, 1999, \apj, 517, 565

\bibitem[Podsiadlowski et al.(2002)]{prp02} Podsiadlowski, 
P., Rappaport, S., \& Pfahl, E.~D.\ 2002, \apj, 565, 1107 

\bibitem[Pfahl et al.(2003)]{prp03} Pfahl, E., Rappaport, S., 
\& Podsiadlowski, P.\ 2003, \apj, 597, 1036 

\bibitem[Paczynski \& Zytkow(1978)]{pz78} Paczynski, B., \& Zytkow, A.~N.\ 1978, \apj, 222, 604 

\bibitem[Plavec et al.(1973)]{pup73} Plavec, M., Ulrich, 
R.~K., \& Polidan, R.~S.\ 1973, \pasp, 85, 769 

\bibitem[Podsiadlowski 
\& Mohamed(2007)]{pm07} Podsiadlowski, P., \& Mohamed, S.\ 2007, Baltic Astronomy, 16, 26 

\bibitem[{{Prialnik} \& {Kovetz}(1995)}]{pk95}
{Prialnik} D., {Kovetz} A., 1995, \apj, 445, 789

\bibitem[{{Rappaport}, {Verbunt} \& {Joss}(1983){Rappaport}, {Verbunt}, \&
  {Joss}}]{rvj83}
{Rappaport} S., {Verbunt} F., {Joss} P.~C., 1983, \apj, 275, 713

\bibitem[{{Ricker} \& {Taam}(2012)}]{rt12}
{Ricker} P.~M., {Taam} R.~E., 2012, \apj, 746, 74

\bibitem[{{Riess} {et~al}\mbox{.}(1998){Riess}, {Filippenko}, {Challis},
  {Clocchiatti}, {Diercks}, {Garnavich}, {Gilliland}, {Hogan}, {Jha},
  {Kirshner}, {Leibundgut}, {Phillips}, {Reiss}, {Schmidt}, {Schommer},
  {Smith}, {Spyromilio}, {Stubbs}, {Suntzeff}, \& {Tonry}}]{rfcc+98}
{Riess} A.~G. {et~al.}, 1998, \aj, 116, 1009

\bibitem[Ritter(1988)]{1988A&A...202...93R} Ritter, H.\ 1988, \aap, 202, 93 

\bibitem[{{Ruiter}, {Belczynski} \& {Fryer}(2009){Ruiter}, {Belczynski}, \&
  {Fryer}}]{rbf09}
{Ruiter} A.~J., {Belczynski} K., {Fryer} C., 2009, \apj, 699, 2026

\bibitem[Toonen et 
al.(2014)]{tcmr13} Toonen, S., Claeys, J.~S.~W., Mennekens, N., \& Ruiter, A.~J.\ 2014, \aap, 562, A14 

\bibitem[{{Totani} {et~al}\mbox{.}(2008){Totani}, {Morokuma}, {Oda}, {Doi}, \&
  {Yasuda}}]{tmod+08}
{Totani} T., {Morokuma} T., {Oda} T., {Doi} M., {Yasuda} N., 2008, \pasj, 60,
  1327

\bibitem[{{Tutukov} \& {Yungelson}(1981)}]{ty81}
{Tutukov} A.~V., {Yungelson} L.~R., 1981, Nauchnye Informatsii, 49, 3

\bibitem[{{Wang}, {Li} \& {Han}(2010){Wang}, {Li}, \& {Han}}]{wlh10}
{Wang} B., {Li} X.-D., {Han} Z.-W., 2010, \mnras, 401, 2729

\bibitem[{{Webbink}(1984)}]{webb84}
{Webbink} R.~F., 1984, \apj, 277, 355

\bibitem[{{Whelan} \& {Iben}(1973)}]{wi73}
{Whelan} J., {Iben}, Jr. I., 1973, \apj, 186, 1007

\bibitem[Willems 
\& Kolb(2004)]{wk04} Willems, B., \& Kolb, U.\ 2004, \aap, 419, 1057 

\bibitem[Wolf et al.(2013)]{wbbp13} Wolf, W.~M., Bildsten, L., 
Brooks, J., \& Paxton, B.\ 2013, \apj, 777, 136 

\bibitem[Woods \& Gilfanov(2013)]{2013MNRAS.432.1640W} Woods, T.~E., \& Gilfanov, M.\ 2013, \mnras, 432, 1640 

\bibitem[Woods \& Gilfanov(2014)]{2014MNRAS.tmp..350W} Woods, T.~E., \& Gilfanov, M.\ 2014, \mnras, 350 

\bibitem[{{Woods} \& {Ivanova}(2011)}]{wi11}
{Woods} T.~E., {Ivanova} N., 2011, \apjl, 739, L48

\bibitem[Xu 
\& Li(2010)]{xl10} Xu, X.-J., \& Li, X.-D.\ 2010, \apj, 716, 114 

\bibitem[{{Yaron} {et~al}\mbox{.}(2005){Yaron}, {Prialnik}, {Shara}, \&
  {Kovetz}}]{ypsk05}
{Yaron} O., {Prialnik} D., {Shara} M.~M., {Kovetz} A., 2005, \apj, 623, 398

\bibitem[{{Yungelson} {et~al}\mbox{.}(1996){Yungelson}, {Livio}, {Truran},
  {Tutukov}, \& {Fedorova}}]{yltt+96}
{Yungelson} L., {Livio} M., {Truran} J.~W., {Tutukov} A., {Fedorova} A., 1996,
  \apj, 466, 890

\bibitem[{{Yungelson} {et~al}\mbox{.}(1995){Yungelson}, {Livio}, {Tutukov}, \&
  {Kenyon}}]{yltk95}
{Yungelson} L., {Livio} M., {Tutukov} A., {Kenyon} S.~J., 1995, \apj, 447, 656

\bibitem[Yungelson 
\& Livio(1998)]{1998ApJ...497..168Y} Yungelson, L., \& Livio, M.\ 1998, \apj, 497, 168

\bibitem[{{Yungelson}(2010)}]{y10}
{Yungelson} L.~R., 2010, Astronomy Letters, 36, 780

\bibitem[{{Zorotovic} {et~al}\mbox{.}(2010){Zorotovic}, {Schreiber},
  {G{\"a}nsicke}, \& {Nebot G{\'o}mez-Mor{\'a}n}}]{zsgn10}
{Zorotovic} M., {Schreiber} M.~R., {G{\"a}nsicke} B.~T., {Nebot
  G{\'o}mez-Mor{\'a}n} A., 2010, \aap, 520, A86

\bibitem[Zorotovic et 
al.(2014)]{zsp14} Zorotovic, M., Schreiber, M.~R., \& Parsons, S.~G.\ 2014, \aap, 568, L9 
\end{thebibliography}


\label{lastpage}

\end{document}